\documentclass{IEEEtran}
\usepackage[latin9]{inputenc}
\usepackage{color}
\usepackage{array}
\usepackage{units}
\usepackage{multirow}
\usepackage{amsmath}
\usepackage{amssymb}
\usepackage{graphicx}
\usepackage{esint}

\makeatletter

\providecommand{\tabularnewline}{\\}

\usepackage{cite}
\usepackage{amsfonts}\usepackage{algorithmic}
\def\BibTeX{{\rm B\kern-.05em{\sc i\kern-.025em b}\kern-.08em
    T\kern-.1667em\lower.7ex\hbox{E}\kern-.125emX}}

\usepackage{caption}

\usepackage{stackengine}

\usepackage{widetext}

\ifdefined\showcaptionsetup
 \PassOptionsToPackage{caption=false}{subfig}
\fi
\usepackage{subfig}
\makeatother

\begin{document}
\title{Surface Plasmon Polaritons: Creation Dynamics and Interference of
Slow and Fast Propagating SPPs at a Temporal Boundary}
\author{Jay A. Berres, \IEEEmembership{Member, IEEE}, S. Ali Hassani Gangaraj,
\IEEEmembership{Member, IEEE}, and George W. Hanson, \IEEEmembership{Fellow, IEEE}
\thanks{Manuscript received ...} \thanks{J. A. Berres is with the Department of Electrical Engineering, University
of Wisconsin-Milwaukee, Milwaukee, WI 53211 USA (e-mail: jaberres@gmail.com). } \thanks{S. A. Hassani Gangaraj is with the Optical Physics Division, Corning
Inc., 184 Science Center Dr., Painted Post, NY 14870, USA (e-mail:
ali.gangaraj@gmail.com).} \thanks{G. W. Hanson is with the Department of Electrical Engineering, University
of Wisconsin-Milwaukee, Milwaukee, WI 53211 USA (e-mail: george@uwm.edu).}}
\maketitle
\begin{abstract}
We establish the theoretical framework for a material system that
supports surface plasmon polaritions (SPPs) excited by a dipole excitation,
where the media configuration suddenly changes at a temporal boundary.
We employ three-dimensional Green's function analysis in the Laplace
transform domain. We use this framework to demonstrate dynamic SPP
formation and time-boundary-induced interference of slow and fast
propagating SPPs. This analysis provides insight into how SPPs are
formed in time and how they interfere at a temporal boundary.
\end{abstract}


\section{Introduction\label{sec:Introduction}}

\IEEEPARstart{E}{lectromagnetic} wave propagation in time-varying
media has long been a topic of research, with early investigations
focusing on the electromagnetic response in media undergoing temporal
changes to the material parameters (i.e., a temporal boundary). These
early investigations considered the propagation of waves within a
spatially homogeneous time-varying dielectric \cite{Morgenthaler1958},
pulsed excitations within a spatially homogeneous time-varying plasma
\cite{Felsen1970}, and a vacuum-dielectric half-space, with a temporal
boundary, where the waves are incident on the dielectric from the
vacuum space \cite{Fante1971}. Furthermore, \cite{Ruiz1978} considered
a spatially homogeneous dielectric with a time-boundary using Laplace
transforms, and, finally, notably, \cite{Jiang1975} considered a
dipole excitation (electric dipole point source) within a spatially
homogeneous dielectric (vacuum), where the dielectric suddenly changes
to a plasma at a temporal boundary, using the Laplace transform technique.

In all of these cases, and subsequent research \cite{Xiao2014,Bakunov2014,Caloz2020a,Caloz2020,Solis2021,Gratus2021,Galiffi2022,Huidobro2023},
it is established that a time-varying media platform results in unique
electromagnetic phenomenon not seen in time-static media, e.g., frequency
shifting and frequency splitting of the incident wave at the temporal
boundary, due to the conservation of momentum, which results in spatially
forward and backward waves (frequency splitting) at different frequencies
than the incident wave (frequency shifting) in the temporal region
after the temporal boundary.

Due to the irreversibility of time, the waves scattered at a temporal
boundary can only occur in the temporal region after the temporal
boundary, i.e., they can't travel back to the past (to the temporal
region before the temporal boundary). This behavior is contrary to
that of wave scattering at a spatial boundary, resulting in reflected
waves and transmitted waves relative to the spatial boundary.

Lately, there has been a resurgence in this research topic \cite{Bakunov2014,Xiao2014,Bacot2016,Sounas2017,Maslov2018,Lustig2018,Shirokova2019,Zhou2020,PachecoPena2020,PachecoPena2020a,Bakunov2020,Ramaccia2020,Solis2021,Ramaccia2021,Xu2021,Sharabi2021,Carminati2021,Gratus2021,Li2021,Li2022,Galiffi2022,VazquezLozano2022,Mencagli2022,Stefanini2022,Koutserimpas,VazquezLozano2023,Wang2023,Koivurova2023,Horsley2023,Mostafa2024,Kreiczer2024,Raziman},
driven by the potential for the applications in plasmonics (nano-photonics,
nano-optics, photonic metamaterials), such as magnet-free nonreciprocity
\cite{Sounas2017,Li2022}, temporal aiming \cite{PachecoPena2020a,Sini2024},
and extreme energy transformations \cite{Li2021}, to name a few.
There has been some work on time-varying media systems that support
SPPs \cite{Maslov2018,Shirokova2019,Bakunov2020,Raziman}; however,
these deal with existing propagating SPPs encountering a temporal
boundary. To the best of our knowledge, there has not been work on
dipole excitation of SPPs in a time-varying system. This is the topic
of this work \cite{Berres2024}, where such a system allows for the
analysis of the interaction among dipole excitations, where we have
the ability to consider the interactions at the moment of SPP creation
(e.g., as a time-varying system changes from one that does not support
SPPs to one that supports SPP propagation). Dipole excitation of SPPs
in inhomogeneous media pre- or post-time-boundary, and the interaction
of before- and after-time-boundary SPPs with continuous excitation,
is the novel aspect of this work.

The use of a time-varying media platform allows for the utilization
of temporal modulation, which, when combined with plasmonic waveguides
(reciprocal or nonreciprocal), may allow for modifying the resonance
or direction of energy propagation in the system. This then may also
allow for more freedom and practicality in manipulating the electromagnetic
response of plasmonic waveguides in time, which in turn may enable
more efficient and tunable interactions among dipole excitations.

The typical approach to study these types of systems is to define
an unbounded homogeneous space, where abrupt changes in time of the
electromagnetic properties of a material in this space create temporal
boundaries that replace the spatial boundaries of a related time-invariant
material configuration. At these temporal boundaries momentum is conserved,
however, the angular frequencies vary in time, i.e., they are specific
to the temporal region before and after the switching event \cite{Xiao2014,Bakunov2014,Gratus2021,Solis2021,Galiffi2022,Li2022},
in order for the dispersion relation to be satisfied for each temporal
region. That is, momentum is conserved at these temporal boundaries
due to the preserved spatial homogeneity, and due to the need for
the temporal boundary conditions to be satisfied everywhere in space.
At the temporal boundary, all components of the electric displacement
$\mathbf{D}$ and the magnetic flux density $\mathbf{B}$ are continuous,
regardless of the medium model (e.g., nondispersive or dispersive),
for any position in space \cite{Xiao2014,Bakunov2014,Gratus2021,Solis2021,Galiffi2022,Li2022}.
In the case of dispersive media \cite{Bakunov2014}, the nonlocal
in time constitutive relations lead to the continuity of all of the
components of $\mathbf{E}$, $\mathbf{H}$, and $\mathbf{J}$ at the
temporal boundary for any position in space. The fields are then matched
at the temporal boundary for all field modes allowed to propagate.
That is, the recipe is to apply momentum conservation at the temporal
boundary and then use the dispersion relation to determine the allowed
modal frequencies. Then, determine the allowed field modes and match
the fields at the temporal boundary to solve for the time-dependent
field amplitudes. One then has the field solutions for each temporal
region at the temporal boundary. This is the approach used in most
cases where the electromagnetic waves can be decomposed (modal expansions
can be performed) to obtain the wavenumbers for the propagating waves
in each temporal region.

However, this method becomes difficult in the case of a time-varying
media system that supports SPPs excited by a dipole excitation (electric
dipole point source), e.g., a single dielectric-plasma interface with
a dipole excitation above the interface, because there are numerous
wavenumbers, in multiple directions, supporting multiple propagating
waves (radiation modes (radiating waves), SPP modes (surface waves),
Brewster (bulk) modes (waves in the plasma (bulk) region)) that satisfy
the propagation constraints of the spatial configuration. Additionally,
for each one of these allowed propagation constants there are numerous
modal temporal frequencies that satisfy the propagation constraints
of the temporal configuration (momentum conservation at the temporal
boundary).

Therefore, here we instead utilize the Laplace transform technique,
which accounts for the temporal boundary conditions in the initial
conditions of the Laplace transform. The inverse Laplace transform
naturally sums up all of the contributions from the allowed propagating
modes for the allowed modal temporal frequencies ($s$-values) for
the allowed modal spatial frequencies ($q$-values).

In the subsequent sections we establish the theoretical framework
for the specific configuration of a time-varying media system that
supports SPPs, in at least one temporal region, or both, excited by
a dipole source, where the media configuration suddenly changes at
a temporal boundary. We first state Maxwell's equations in the Laplace
transform domain, derive the inhomogeneous electric field wave equation
in the Laplace transform domain for generic media, i.e., linear, inhomogeneous,
anisotropic, temporally dispersive (temporally nonlocal), and spatially
nondispersive (local) media, and then illustrate the use of Green's
functions to solve the inhomogeneous wave equation to obtain the electric
field.

We then consider the special case of isotropic media and one temporal
boundary, where we derive the corresponding Green's functions and
demonstrate dynamic SPP formation and interference of slow and fast
propagating SPPs at the temporal boundary.

\begin{figure}
\begin{centering}
\includegraphics[width=1\columnwidth]{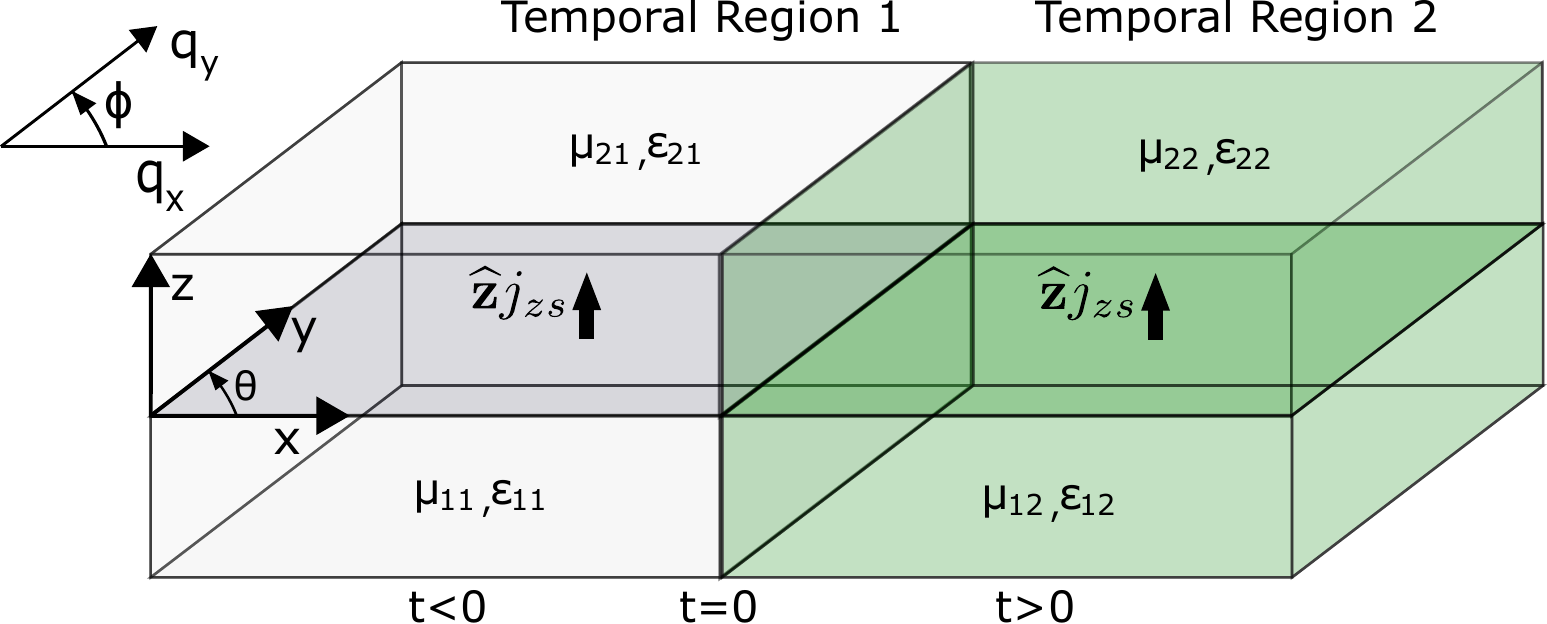}
\par\end{centering}
\caption{Model for the general case of a single interface between two different
materials ($\mu_{2}$,$\varepsilon_{2}$ for $z>0$, $\mu_{1}$,$\varepsilon_{1}$
for $z<0$) with an electric dipole point source, $\widehat{\mathbf{z}}j_{zs}$,
located along the interface surface, where we consider two temporal
regions (one temporal boundary). The double subscript notation is
described in the text. Additionally, $q_{x}$, $q_{y}$, and $\phi$
define the coordinate system in the momentum space. The material configurations
investigated in this work, as established in Tables \ref{tab:Relevant-values-used}
and \ref{tab:Relevant-values-used-1}, consist of air and plasma,
where the spatial regions are assumed to be piece-wise homogeneous.
\label{fig:Model-for-the-1}}
\end{figure}

The model for this configuration, seen in Fig. \ref{fig:Model-for-the-1},
defines two temporal regions (one temporal boundary). This framework
can be extended to the multi-temporal boundary case by solving the
inhomogeneous electric field wave equation in the Laplace transform
domain in each temporal region, using the Green's functions applicable
to each temporal region, where the initial conditions at the preceding
temporal boundary are applicable. 
 We use dispersive materials to model realistic systems, and which
ensure that the temporal boundary conditions are met for $\mathbf{E}$,
$\mathbf{H}$, and $\mathbf{J}$ (i.e., we have a temporally causal
system).

Note that in this work we simply consider that the media parameters
are suddenly changed. In practice, this can be achieved in material
systems with free electrons, such as plasmas, semiconductors, and
two-dimensional systems such as graphene, by the change in a DC bias
(DC magnetic field or DC voltage). Although those systems are anisotropic,
here we consider isotropic materials for simplicity; the extension
to anisotropic systems is straightforward as the material configuration
is described by the Green's function.

\section{Theoretical Model\label{sec:Theoretical-Model}}

\subsection{Maxwell's Equations in the Laplace Transform Domain}

Maxwell's equations in matter, in the Laplace transform domain, can
be written as \cite{Zangwill2013,Lathi2005}
\begin{align}
\mathbf{\boldsymbol{\nabla}}\times\mathbf{E}(\mathbf{r},s) & =-s\mu_{0}\underline{\boldsymbol{\mu}}_{r}(\mathbf{r},s)\cdot\mathbf{H}(\mathbf{r},s)\label{eq:Faraday's law_Laplace-2}\\
 & +\mu_{0}\mathbf{H}(\mathbf{r},t^{-})+\mu_{0}\mathbf{M}(\mathbf{r},t^{-}),\nonumber \\
\mathbf{\boldsymbol{\nabla}}\times\mathbf{H}(\mathbf{r},s) & =s\varepsilon_{0}\underline{\boldsymbol{\varepsilon}}_{r}(\mathbf{r},s)\cdot\mathbf{E}(\mathbf{r},s)\label{eq:Ampere's law_Laplace-2}\\
 & -\varepsilon_{0}\mathbf{E}(\mathbf{r},t^{-})-\mathbf{P}(\mathbf{r},t^{-})+\mathbf{j}_{s}(\mathbf{r},s),\nonumber \\
\mathbf{\boldsymbol{\nabla}}\cdot\mathbf{D}(\mathbf{r},s) & =\rho_{s}(\mathbf{r},s),\label{eq:Gauss' law_Laplace-2}\\
\mathbf{\boldsymbol{\nabla}}\cdot\mathbf{B}(\mathbf{r},s) & =0,\label{eq:magnetic Gauss' law_Laplace-2}
\end{align}
where $\mathbf{E}$ is the electric field intensity, $\mathbf{B}$
is the magnetic flux density, $\rho$ is the source charge density,
$\mathbf{j}$ is the source current density, $\varepsilon_{0}$ is
the permittivity of vacuum, $\mu_{0}$ is the permeability of vacuum,
$\underline{\boldsymbol{\varepsilon}}_{r}(\mathbf{r},s)$ is the relative
permittivity tensor, $\underline{\boldsymbol{\mu}}_{r}(\mathbf{r},s)$
is the relative permeability tensor, $\mathbf{D}$ is the electric
flux density, $\mathbf{H}$ is the magnetic field intensity, $\mathbf{P}$
is the polarization, and $\mathbf{M}$ is the magnetization. Additionally,
variables that are a function of $t^{-}$ are the initial conditions
for that variable at the time just before time $t$.

\subsection{Inhomogeneous Wave Equation for the Electric Field in the Laplace
Transform Domain}

We now derive the inhomogeneous wave equation for the electric field
in the Laplace transform domain. We can then solve for the electric
field in each medium. Assuming non-magnetic media, $\underline{\boldsymbol{\mu}}_{r}(\mathbf{r},s)=\underline{\mathbf{I}}$
and $\mathbf{M}(\mathbf{r},t^{-})=\mathbf{0}$, multiplying both sides
of (\ref{eq:Faraday's law_Laplace-2}) by $\mu_{\mathrm{0}}^{-1}\underline{\boldsymbol{\mu}}^{-1}(\mathbf{r},s)$,
taking the curl of both sides, inserting (\ref{eq:Ampere's law_Laplace-2}),
and multiplying both sides by $\mu_{0}$, we have
\begin{align}
\mathbf{\boldsymbol{\nabla}}\times\mathbf{\boldsymbol{\nabla}}\times\mathbf{E}(\mathbf{r},s) & =-\frac{s^{2}}{c^{2}}\underline{\boldsymbol{\varepsilon}}_{r}(\mathbf{r},s)\cdot\mathbf{E}(\mathbf{r},s)\label{eq:E_Wave Equation_Laplace-1-1-1-1-2-1}\\
 & -s\mu_{0}\mathbf{j}_{s}(\mathbf{r},s)+\frac{s}{c^{2}}\mathbf{E}(\mathbf{r},t^{-})\nonumber \\
 & +\mu_{0}s\mathbf{P}(\mathbf{r},t^{-})+\mu_{0}\mathbf{\boldsymbol{\nabla}}\times\mathbf{H}(\mathbf{r},t^{-}),\nonumber 
\end{align}
where we used $\nicefrac{1}{c^{2}}=\mu_{0}\varepsilon_{0}$. Equation
(\ref{eq:E_Wave Equation_Laplace-1-1-1-1-2-1}) is the inhomogeneous
wave equation for the electric field in the Laplace transform domain.

In order to obtain a more usable form, where we can incorporate common
initial conditions, using Ampere's law in the time domain, $\mathbf{\boldsymbol{\nabla}}\times\mathbf{H}(\mathbf{r},t)=\frac{\partial}{\partial t}\mathbf{D}(\mathbf{r},t)+\mathbf{j}_{s}(\mathbf{r},t)\rightarrow\mathbf{\boldsymbol{\nabla}}\times\mathbf{H}(\mathbf{r},t^{-})=\frac{\partial}{\partial t}\mathbf{D}(\mathbf{r},t)\left|_{t=t^{-}}\right.+\mathbf{j}_{s}(\mathbf{r},t^{-})$,
and the electric constitutive relation in the time domain, $\mathbf{D}(\mathbf{r},t)=\varepsilon_{0}\mathbf{E}(\mathbf{r},t)+\mathbf{P}(\mathbf{r},t)\rightarrow\mathbf{P}(\mathbf{r},t^{-})=\mathbf{D}(\mathbf{r},t^{-})-\varepsilon_{0}\mathbf{E}(\mathbf{r},t^{-})$,
we can write (\ref{eq:E_Wave Equation_Laplace-1-1-1-1-2-1}) as
\begin{align}
\mathbf{\boldsymbol{\nabla}}\times\mathbf{\boldsymbol{\nabla}}\times\mathbf{E}(\mathbf{r},s) & =-\frac{s^{2}}{c^{2}}\underline{\boldsymbol{\varepsilon}}_{r}(\mathbf{r},s)\cdot\mathbf{E}(\mathbf{r},s)\label{eq:E_Wave Equation_Laplace-1-1-1-1-2-1-1}\\
 & -s\mu_{0}\mathbf{j}_{s}(\mathbf{r},s)+\mu_{0}\mathbf{j}_{s}(\mathbf{r},t)\left|_{t=t^{-}}\right.\nonumber \\
 & +\mu_{0}\frac{\partial}{\partial t}\mathbf{D}(\mathbf{r},t)\left|_{t=t^{-}}\right.+\mu_{0}s\mathbf{D}(\mathbf{r},t)\left|_{t=t^{-}}\right..\nonumber 
\end{align}

We can use (\ref{eq:E_Wave Equation_Laplace-1-1-1-1-2-1-1}) to solve
for the electric field in any temporal region, where the initial conditions
are known at the temporal boundary immediately preceding the temporal
region. We assume that there is no change in the induced current density
just after the time-change, i.e., at $t=t_{\alpha}^{+}$, where $\alpha=0,1,2,...$.
That is, although the material permittivity is modeled as changing
value instantaneously, the material response (electrons) cannot change
instantaneously due to their mass. This leads to the temporal boundary
conditions (BCs); all components of the electric field $\mathbf{E}$,
the magnetic field $\mathbf{H}$, and the current density $\mathbf{J}$
are continuous at the temporal boundary. Then the fields are the same
just before and just after the time-change with respect to the media
(and consequently at the time-change $t=t_{\alpha}$), i.e., at $t=t_{\alpha}^{-}$
and $t=t_{\alpha}^{+}$. In other words, we can use $t=t_{\alpha}^{-}$,
$t=t_{\alpha}$, or $t=t_{\alpha}^{+}$ for the initial condition
parameters.

We can use (\ref{eq:E_Wave Equation_Laplace-1-1-1-1-2-1-1}) as is
for a known initial condition in the time domain, which we will set
for the case of temporal region 1 since we set the initial conditions
that start the system. Once we solve for the fields in temporal region
1 we will have the initial conditions to solve for the fields in temporal
region 2.

\subsection{One Temporal Boundary}

Here, we consider the case of one temporal boundary at $t=t_{0}=0$.
We assume the sources and fields in temporal region 1 are time-harmonic
(monochromatic, sinusoidal steady-state) with time variations of the
form $e^{-i\omega_{0}t}$. At this point we formally adopt the notation
of $x_{mn}$, where $x$ is some parameter, where $m=1,2,3,...$ designates
its spatial region and $n=1,2,3,...$ designates its temporal region.
We then write (\ref{eq:E_Wave Equation_Laplace-1-1-1-1-2-1-1}) for
temporal region 2 as
\begin{align}
\mathbf{\boldsymbol{\nabla}}\times\mathbf{\boldsymbol{\nabla}}\times\mathbf{E}_{m2}(\mathbf{r},s) & =-\frac{s^{2}}{c^{2}}\underline{\boldsymbol{\varepsilon}}_{rm2}(\mathbf{r},s)\cdot\mathbf{E}_{m2}(\mathbf{r},s)\label{eq:E_Wave Equation_Laplace-1-1-1-1-2-1-1-1}\\
 & -s\mu_{0}\mathbf{j}_{s22}(\mathbf{r},s)+\mu_{0}\mathbf{j}_{s21}(\mathbf{r},t)\left|_{t_{0}=0}\right.\nonumber \\
 & +\mu_{0}\frac{\partial}{\partial t}\mathbf{D}_{m1}(\mathbf{r},t)\left|_{t_{0}=0}\right.\nonumber \\
 & +\mu_{0}s\mathbf{D}_{m1}(\mathbf{r},t)\left|_{t_{0}=0}\right.,\nonumber 
\end{align}
where $\mathbf{j}_{s21}(\mathbf{r},t)=\mathrm{Re}\left\{ \mathbf{j}_{s21}(\mathbf{r})e^{-i\omega_{0}t}\right\} \rightarrow\mathbf{j}_{s21}(\mathbf{r})e^{-i\omega_{0}t}$
and $\mathbf{D}_{m1}(\mathbf{r},t)=\mathrm{Re}\left\{ \mathbf{D}_{m1}(\mathbf{r})e^{-i\omega_{0}t}\right\} \rightarrow\mathbf{D}_{m1}(\mathbf{r})e^{-i\omega_{0}t}$.\footnote{We drop the $\mathrm{Re}\left\{ \cdot\right\} $ notation to simplify
calculations, where we will then take the $\mathrm{Re}\left\{ \cdot\right\} $
of the final expressions in the time domain to ensure that the values
are real in the time domain. We note that this only applies to the
expressions in temporal region 1, where we assumed time-harmonic conditions.} We assume that the source current density remains the same for all
of time in all temporal regions, then we can write $\mathbf{j}_{s21}(\mathbf{r},t)\rightarrow\mathbf{j}_{s}(\mathbf{r})e^{-i\omega_{0}t}$
and $\mathbf{j}_{s22}(\mathbf{r},s)=\mathbf{j}_{s21}(\mathbf{r},s)=\mathcal{L}\left\{ \mathbf{j}_{s21}(\mathbf{r},t)\right\} =\mathcal{L}\left\{ \mathbf{j}_{s}(\mathbf{r})e^{-i\omega_{0}t}\right\} =\frac{1}{s+i\omega_{0}}\mathbf{j}_{s}(\mathbf{r})$.
Additionally, with $\mathbf{D}(\mathbf{r})=\varepsilon_{0}\underline{\boldsymbol{\varepsilon}}_{r}(\mathbf{r},\omega_{0})\cdot\mathbf{E}(\mathbf{r})$,
we can write $\mathbf{D}_{m1}(\mathbf{r})=\varepsilon_{0}\underline{\boldsymbol{\varepsilon}}_{rm1}(\mathbf{r},\omega_{0})\cdot\mathbf{E}_{m1}(\mathbf{r})$.
We can then can write (\ref{eq:E_Wave Equation_Laplace-1-1-1-1-2-1-1-1})
as
\begin{align}
\mathbf{\boldsymbol{\nabla}}\times\mathbf{\boldsymbol{\nabla}}\times\mathbf{E}_{m2}(\mathbf{r},s) & =-\frac{s^{2}}{c^{2}}\underline{\boldsymbol{\varepsilon}}_{rm2}(\mathbf{r},s)\cdot\mathbf{E}_{m2}(\mathbf{r},s)\label{eq:E_Wave Equation_Laplace-1-1-1-1-2-1-1-1-1-1}\\
 & +\frac{i\omega_{0}\mu_{0}}{s+i\omega_{0}}\mathbf{j}_{s}(\mathbf{r})\nonumber \\
 & +\frac{s-i\omega_{0}}{c^{2}}\underline{\boldsymbol{\varepsilon}}_{rm1}(\mathbf{r},\omega_{0})\cdot\mathbf{E}_{m1}(\mathbf{r}),\nonumber 
\end{align}
which is the inhomogeneous wave equation for the electric field in
the Laplace transform domain (when used for spatial region 1 we set
$\mathbf{j}_{s}(\mathbf{r})$ to zero).

We can use the inhomogeneous wave equation for the electric field
in the spatial regions in temporal region 1,
\begin{align}
\mathbf{\boldsymbol{\nabla}}\times\mathbf{\boldsymbol{\nabla}}\times\mathbf{E}_{m1}(\mathbf{r}) & =\frac{\omega_{0}^{2}}{c^{2}}\underline{\boldsymbol{\varepsilon}}_{rm1}(\mathbf{r},\omega_{0})\cdot\mathbf{E}_{m1}(\mathbf{r})\label{eq:E_Wave Equation_Laplace-1-1-1-1-1-1}\\
 & +i\omega_{0}\mu_{0}\mathbf{j}_{s}(\mathbf{r}),\nonumber 
\end{align}
to solve for the electric field in the spatial regions in temporal
region 1 (when used for spatial region 1 we set $\mathbf{j}_{s}(\mathbf{r})$
to zero). We can solve for the magnetic field in temporal region 1
using
\begin{align}
\mathbf{H}_{m1}(\mathbf{r}) & =\frac{1}{i\omega_{0}\mu_{0}}\mathbf{\boldsymbol{\nabla}}\times\mathbf{E}_{m1}(\mathbf{r}).\label{eq:magnetic field in temporal region 1}
\end{align}

\subsection{Obtaining the Standard Form for the Wave Equation}

The solution for the fields in temporal region 1 follows any usual
(non-time-varying media) method, such as using Green's functions,
which then leads to the field values for the temporal BCs. For temporal
region 2 we also use the Green's function, although we first need
to cast (\ref{eq:E_Wave Equation_Laplace-1-1-1-1-2-1-1-1-1-1}) into
a standard form (from which the Green's function follows in a straightforward
manner). In (\ref{eq:E_Wave Equation_Laplace-1-1-1-1-2-1-1-1-1-1})
we can see that we have an extra term on the right hand side of the
wave equation (a source-like term arising from the initial condition),
which results in the wave equation not being in the standard form.
Following \cite{Jiang1975}, we can obtain the standard form for the
wave equation by performing a transformation regarding this term.
We start by rewriting $\mathbf{E}_{m1}(\mathbf{r})$ as
\begin{align}
\mathbf{E}_{m1}(\mathbf{r}) & =\mathbf{E}_{m1}(\mathbf{r})\cdot\left(\underline{\mathbf{F}}_{1}(\mathbf{r},s)+\underline{\mathbf{F}}_{2}(\mathbf{r},s)\right),\label{eq:F_coefficients-2-3}
\end{align}
where we assume $\underline{\mathbf{F}}_{1}$ and $\underline{\mathbf{F}}_{2}$
are tensors, functions of $\mathbf{r}$ and $s$, and they satisfy
the unity relation $\underline{\mathbf{F}}_{1}(\mathbf{r},s)+\underline{\mathbf{F}}_{2}(\mathbf{r},s)=\underline{\mathbf{I}}$.
Using (\ref{eq:E_Wave Equation_Laplace-1-1-1-1-1-1}), we can write
\begin{align}
\mathbf{E}_{m1}(\mathbf{r}) & =\frac{\underline{\boldsymbol{\varepsilon}}_{rm1}^{-1}(\mathbf{r},\omega_{0})}{k_{0}^{2}}\cdot\left(\mathbf{\boldsymbol{\nabla}}\times\mathbf{\boldsymbol{\nabla}}\times\mathbf{E}_{m1}(\mathbf{r})-i\omega_{0}\mu_{0}\mathbf{j}_{s}(\mathbf{r})\right),\label{eq:temporal region 1_wave equation_general-1-3}
\end{align}
where we note that $k_{0}=\nicefrac{\omega_{0}}{c}=\omega_{0}\sqrt{\mu_{0}\varepsilon_{0}}$.
Then (\ref{eq:F_coefficients-2-3}) can be written as
\begin{align}
\mathbf{E}_{m1}(\mathbf{r}) & =\mathbf{E}_{m1}(\mathbf{r})\cdot\underline{\mathbf{F}}_{1}(\mathbf{r},s)\label{eq:F_coefficients_1-1-3}\\
 & +\frac{\underline{\boldsymbol{\varepsilon}}_{rm1}^{-1}(\mathbf{r},\omega_{0})}{k_{0}^{2}}\cdot\mathbf{\boldsymbol{\nabla}}\times\mathbf{\boldsymbol{\nabla}}\times\mathbf{E}_{m1}(\mathbf{r})\cdot\underline{\mathbf{F}}_{2}(\mathbf{r},s)\nonumber \\
 & -\frac{\underline{\boldsymbol{\varepsilon}}_{rm1}^{-1}(\mathbf{r},\omega_{0})}{k_{0}^{2}}\cdot i\omega_{0}\mu_{0}\mathbf{j}_{s}(\mathbf{r})\cdot\underline{\mathbf{F}}_{2}(\mathbf{r},s).\nonumber 
\end{align}
\begin{widetext}Then, after substituting (\ref{eq:F_coefficients_1-1-3})
into (\ref{eq:E_Wave Equation_Laplace-1-1-1-1-2-1-1-1-1-1}) and using
the resulting equation and the unity relation to solve for $\underline{\mathbf{F}}_{1}$
and $\underline{\mathbf{F}}_{2}$, we can write
\begin{align}
\mathbf{\boldsymbol{\nabla}}\times\mathbf{\boldsymbol{\nabla}}\times\mathbf{E}_{m2}^{\prime}(\mathbf{r},s) & =-\frac{s^{2}}{c^{2}}\underline{\boldsymbol{\varepsilon}}_{rm2}(\mathbf{r},s)\cdot\mathbf{E}_{m2}^{\prime}(\mathbf{r},s)-s\mu_{0}\mathbf{j}_{s}^{\prime}(\mathbf{r},s),\label{eq:temporal region 2_wave equation_general_standard form-1-3}
\end{align}
where
\begin{align}
\mathbf{E}_{m2}^{\prime}(\mathbf{r},s) & \equiv\mathbf{E}_{m2}(\mathbf{r},s)-\left(s-i\omega_{0}\right)\left(s^{2}\underline{\boldsymbol{\varepsilon}}_{rm1}^{-1}(\mathbf{r},\omega_{0})\cdot\underline{\boldsymbol{\varepsilon}}_{rm2}(\mathbf{r},s)+\omega_{0}^{2}\underline{\mathbf{I}}\right)^{-1}\cdot\mathbf{E}_{m1}(\mathbf{r}),\label{eq:prime electric field in temporal region 2}\\
\mathbf{j}_{s}^{\prime}(\mathbf{r},s) & \equiv\left(\frac{-i\omega_{0}}{s\left(s+i\omega_{0}\right)}-\frac{-i\omega_{0}}{s}\left(s-i\omega_{0}\right)\left(s^{2}\underline{\boldsymbol{\varepsilon}}_{rm1}^{-1}(\mathbf{r},\omega_{0})\cdot\underline{\boldsymbol{\varepsilon}}_{rm2}(\mathbf{r},s)+\omega_{0}^{2}\underline{\mathbf{I}}\right)^{-1}\right)\cdot\mathbf{j}_{s}(\mathbf{r}).\label{eq:current density temporal region 2}
\end{align}
Equation (\ref{eq:temporal region 2_wave equation_general_standard form-1-3})
is now in the standard form for a wave equation, which aligns with
(\ref{eq:E_Wave Equation_Laplace-1-1-1-1-1-1}).

We can determine the magnetic field, with $\mathbf{H}_{m1}(\mathbf{r},t^{-})\rightarrow\mathbf{H}_{m1}(\mathbf{r},t)\left|_{t_{0}=0}\right.$,
where $\mathbf{H}_{m1}(\mathbf{r},t)=\mathrm{Re}\left\{ \mathbf{H}_{m1}(\mathbf{r})e^{-i\omega_{0}t}\right\} \rightarrow\mathbf{H}_{m1}(\mathbf{r})e^{-i\omega_{0}t}$,
as
\begin{align}
\mathbf{H}_{m2}(\mathbf{r},s) & =\frac{1}{-s\mu_{0}}\mathbf{\boldsymbol{\nabla}}\times\mathbf{E}_{m2}(\mathbf{r},s)+\frac{1}{i\omega_{0}\mu_{0}s}\mathbf{\boldsymbol{\nabla}}\times\mathbf{E}_{m1}(\mathbf{r}).\label{eq:magnetic field_temporal region 2}
\end{align}
Using (\ref{eq:prime electric field in temporal region 2}) we can
write an expression for $\mathbf{E}_{m2}(\mathbf{r},s)$ and substitute
it into (\ref{eq:magnetic field_temporal region 2}) to obtain
\begin{align}
\mathbf{H}_{m2}(\mathbf{r},s) & =\mathbf{H}_{m2}^{\prime}(\mathbf{r},s)+\frac{1}{-s\mu_{0}}\mathbf{\boldsymbol{\nabla}}\times\left(\left(s-i\omega_{0}\right)\left(s^{2}\underline{\boldsymbol{\varepsilon}}_{rm1}^{-1}(\mathbf{r},\omega_{0})\cdot\underline{\boldsymbol{\varepsilon}}_{rm2}(\mathbf{r},s)+\omega_{0}^{2}\underline{\mathbf{I}}\right)^{-1}\cdot\mathbf{E}_{m1}(\mathbf{r})\right)+\frac{1}{i\omega_{0}\mu_{0}s}\mathbf{\boldsymbol{\nabla}}\times\mathbf{E}_{m1}(\mathbf{r}),\label{eq:magnetic field in temporal region 1-1}
\end{align}
\end{widetext}where $\mathbf{H}_{m2}^{\prime}(\mathbf{r},s)=\frac{1}{-s\mu_{0}}\mathbf{\boldsymbol{\nabla}}\times\mathbf{E}_{m2}^{\prime}(\mathbf{r},s)$.

In the case of piece-wise homogeneous layers, $\underline{\boldsymbol{\varepsilon}}_{rmn}(\mathbf{r})\rightarrow\underline{\boldsymbol{\varepsilon}}_{rmn}$,
and isotropic media, $\underline{\boldsymbol{\varepsilon}}_{rmn}\rightarrow\underline{\mathbf{I}}\varepsilon_{rmn}$,
we can write $\mathbf{E}_{m2}(\mathbf{r},s)=\mathbf{E}_{m2}^{\prime}(\mathbf{r},s)+A_{m2}^{e}(s)\mathbf{E}_{m1}(\mathbf{r})$,
$\mathbf{H}_{m2}(\mathbf{r},s)=\mathbf{H}_{m2}^{\prime}(\mathbf{r},s)+A_{m2}^{m}(s)\mathbf{H}_{m1}(\mathbf{r})$,
and $\mathbf{j}_{s}^{\prime}(\mathbf{r},s)=A_{22}^{s}(s)\mathbf{j}_{s}(\mathbf{r})$,
where
\begin{align}
A_{m2}^{e}(s) & =\frac{\left(s-i\omega_{0}\right)\varepsilon_{rm1}(\omega_{0})}{s^{2}\varepsilon_{rm2}(s)+\omega_{0}^{2}\varepsilon_{rm1}(\omega_{0})},\\
A_{m2}^{m}(s) & =\frac{s\varepsilon_{rm2}(s)-i\omega_{0}\varepsilon_{rm1}(\omega_{0})}{s^{2}\varepsilon_{rm2}(s)+\omega_{0}^{2}\varepsilon_{rm1}(\omega_{0})},
\end{align}
and
\begin{align}
A_{22}^{s}(s)=A_{2}^{s}(s) & =\frac{-i\omega_{0}s^{2}\left(\varepsilon_{r22}(s)-\varepsilon_{r21}(\omega_{0})\right)}{s\left(s+i\omega_{0}\right)\left(s^{2}\varepsilon_{r22}(s)+\omega_{0}^{2}\varepsilon_{r21}(\omega_{0})\right)}.
\end{align}

\subsection{Green's Functions in the Laplace Transform Domain}

We can solve (\ref{eq:temporal region 2_wave equation_general_standard form-1-3})
using the Green's tensor that satisfies
\begin{align}
\left(\mathbf{\boldsymbol{\nabla}}\times\mathbf{\boldsymbol{\nabla}}\times+\frac{s^{2}}{c^{2}}\underline{\boldsymbol{\varepsilon}}_{rm2}(\mathbf{r},s)\cdot\right)\underline{\mathbf{G}}_{m2}^{\prime}(\mathbf{r},\mathbf{r}^{\prime},s) & =\underline{\mathbf{I}}\delta(\mathbf{r}-\mathbf{r}^{\prime}),\label{eq:Green tensor wave equation-1-1-1-1}
\end{align}
where $\underline{\mathbf{G}}_{m2}^{\prime}(\mathbf{r},\mathbf{r}^{\prime},s)$
is the electric field Green's function tensor for the primed electric
field in the Laplace transform domain.

\subsubsection{Spatial Boundary Conditions for the Fields\label{subsec:Spatial-Boundary-Conditions}}

We need to enforce the spatial boundary conditions for the primed-fields
previously defined at the interface. The well established spatial
boundary conditions for the fields at an interface between two media,
where there are no charges or sources along the interface (i.e., the
two media are not perfect conductors and there are no impressed sources
placed along the interface), are \cite{Balanis2009}
\begin{align}
E_{x2n}-E_{x1n}=0, & \;E_{y2n}-E_{y1n}=0,\\
H_{x2n}-H_{x1n}=0, & \;H_{y2n}-H_{y1n}=0.
\end{align}
In the case of temporal region 2, from (\ref{eq:prime electric field in temporal region 2})
and (\ref{eq:magnetic field in temporal region 1-1}) the spatial
boundary conditions on $\mathbf{E}_{m2}(\mathbf{r},s)$ and $\mathbf{H}_{m2}(\mathbf{r},s)$
translate into spatial boundary conditions on $\mathbf{E}_{m2}^{\prime}(\mathbf{r},s)$
and $\mathbf{H}_{m2}^{\prime}(\mathbf{r},s)$, where, for piece-wise
homogeneous layers and isotropic media, and $\beta=x,y$, they are
\begin{align}
E_{\beta22}^{\prime}-E_{\beta12}^{\prime} & =A_{12}^{e}(s)E_{\beta11}-A_{22}^{e}(s)E_{\beta21},\label{eq:prime spatial boundary conditions_electric field}\\
H_{\beta22}^{\prime}-H_{\beta12}^{\prime} & =A_{12}^{m}(s)H_{\beta11}-A_{22}^{m}(s)H_{\beta21}.\label{eq:prime spatial boundary conditions_magnetic field}
\end{align}

\subsubsection{Isotropic Media}

We now establish the Green's function solutions that satisfy (\ref{eq:Green tensor wave equation-1-1-1-1}),
consistent with (\ref{eq:prime spatial boundary conditions_electric field})
and (\ref{eq:prime spatial boundary conditions_magnetic field}),
for the case of piece-wise homogeneous layers and isotropic media.
This framework can be extended to obtain solutions for the case of
anisotropic media by following well-known methods for obtaining Green's
functions for more complicated materials. We derive the Green's functions
and the corresponding electric field solutions for the complete structure
(Fig. \ref{fig:Model-for-the-1}) utilizing the Hertz potentials \cite{Ishimaru1991,Chew1995},
where we define a vertical point dipole current source as a unit point
current source $\mathbf{j}_{s}(\mathbf{r},t)=\widehat{\mathbf{z}}\delta(\mathbf{r}-\mathbf{r}_{0})\cos\left(-\omega_{0}t\right)=\mathrm{Re}\left\{ \mathbf{j}_{s}(\mathbf{r})e^{-i\omega_{0}t}\right\} $,
where $\mathbf{j}_{s}(\mathbf{r})=\widehat{\mathbf{z}}\delta(\mathbf{r}-\mathbf{r}_{0})$.
We note that the Green's functions here for temporal region 2 are
new, since the $\mathbf{E}^{\prime}$ and $\mathbf{H}^{\prime}$ fields
satisfy different spatial boundary conditions than the usual $\mathbf{E}$
and $\mathbf{H}$ fields.

The Hertz potential Green's function components are
\begin{align}
g_{zz2n}^{p}(\mathbf{r},\mathbf{r}^{\prime}) & =\frac{1}{\left(2\pi\right)^{2}}\int_{-\infty}^{\infty}\int_{-\infty}^{\infty}dq_{x}dq_{y}\label{eq:principal Green's function component}\\
 & \frac{e^{-p_{2n}\left|z-z^{\prime}\right|}}{2p_{2n}}e^{i\left(q_{x}\left(x-x^{\prime}\right)+q_{y}\left(y-y^{\prime}\right)\right)}\nonumber \\
 & =\frac{e^{ik_{2n}\sqrt{\rho^{2}+\left(z-z^{\prime}\right)^{2}}}}{4\pi\sqrt{\rho^{2}+\left(z-z^{\prime}\right)^{2}}},\nonumber 
\end{align}
\begin{align}
g_{zz2n}^{s}(\mathbf{r},\mathbf{r}^{\prime}) & =\frac{1}{\left(2\pi\right)^{2}}\int_{-\infty}^{\infty}\int_{-\infty}^{\infty}dq_{x}dq_{y}\label{eq:scattered Green's function component}\\
 & R_{2n}e^{-p_{2n}z}\frac{e^{-p_{2n}z^{\prime}}}{2p_{2n}}e^{i\left(q_{x}\left(x-x^{\prime}\right)+q_{y}\left(y-y^{\prime}\right)\right)},\nonumber 
\end{align}
\begin{align}
g_{zz1n}(\mathbf{r},\mathbf{r}^{\prime}) & =\frac{1}{\left(2\pi\right)^{2}}\int_{-\infty}^{\infty}\int_{-\infty}^{\infty}dq_{x}dq_{y}\\
 & T_{1n}e^{p_{1n}z}\frac{e^{-p_{2n}z^{\prime}}}{2p_{2n}}e^{i\left(q_{x}\left(x-x^{\prime}\right)+q_{y}\left(y-y^{\prime}\right)\right)},\nonumber 
\end{align}
where we can write $g_{zz2n}(\mathbf{r},\mathbf{r}^{\prime})=g_{zz2n}^{p}(\mathbf{r},\mathbf{r}^{\prime})+g_{zz2n}^{s}(\mathbf{r},\mathbf{r}^{\prime})$,
\begin{widetext}with the $p$ superscript designating the principal
field and the $s$ superscript designating the scattered field, $\rho=\sqrt{\left(x-x^{\prime}\right)^{2}+\left(y-y^{\prime}\right)^{2}}$,
and where
\begin{align}
R_{21}=\frac{N_{21}}{D_{21}}, & \;T_{11}=\frac{2p_{21}}{D_{21}},
\end{align}
and

\begin{align}
R_{22}(z^{\prime},s) & =\frac{N_{22}}{D_{22}}\label{eq:scattering coefficient temporal region 2}\\
+\frac{1}{\frac{e^{-p_{22}z^{\prime}}}{2p_{22}}} & \frac{2p_{21}\varepsilon_{r22}(s)\left(\varepsilon_{r12}(s)\varepsilon_{r21}(\omega_{0})-\varepsilon_{r11}(\omega_{0})\varepsilon_{r22}(s)\right)}{\omega_{0}^{2}D_{21}\left(\varepsilon_{r21}(\omega_{0})-\varepsilon_{r22}(s)\right)}\frac{\left(s^{2}+\omega_{0}^{2}\right)\left(s^{2}p_{11}\varepsilon_{r12}(s)+\omega_{0}^{2}p_{12}\varepsilon_{r11}(\omega_{0})\right)}{D_{22}\left(s^{2}\varepsilon_{r12}(s)+\omega_{0}^{2}\varepsilon_{r11}(\omega_{0})\right)}\frac{e^{-p_{21}z^{\prime}}}{2p_{21}},\nonumber \\
T_{12}(z^{\prime},s) & =\frac{2p_{22}}{D_{22}}\\
+\frac{1}{\frac{e^{-p_{22}z^{\prime}}}{2p_{22}}} & \frac{2p_{21}\varepsilon_{r22}(s)\left(\varepsilon_{r12}(s)\varepsilon_{r21}(\omega_{0})-\varepsilon_{r11}(\omega_{0})\varepsilon_{r22}(s)\right)}{\omega_{0}^{2}D_{21}\left(\varepsilon_{r21}(\omega_{0})-\varepsilon_{r22}(s)\right)}\frac{\left(s^{2}+\omega_{0}^{2}\right)\left(s^{2}p_{11}\varepsilon_{r22}(s)-\omega_{0}^{2}p_{22}\varepsilon_{r11}(\omega_{0})\right)}{D_{22}\left(s^{2}\varepsilon_{r12}(s)+\omega_{0}^{2}\varepsilon_{r11}(\omega_{0})\right)}\frac{e^{-p_{21}z^{\prime}}}{2p_{21}},\nonumber 
\end{align}
which are the scattering (reflection) coefficient and the transmission
coefficient, respectively, where $N_{2n}=p_{2n}\varepsilon_{r1n}-p_{1n}\varepsilon_{r2n}$,
$D_{2n}=p_{2n}\varepsilon_{r1n}+p_{1n}\varepsilon_{r2n}$, $q=\sqrt{q_{x}^{2}+q_{y}^{2}}$,
$p_{m2}=\sqrt{q^{2}-k_{m2}^{2}(s)}$, $k_{m2}(s)=is\sqrt{\mu_{0}\varepsilon_{m2}(s)}$,
$\varepsilon_{m2}(s)=\varepsilon_{0}\varepsilon_{rm2}(s)$ and $p_{m1}=\sqrt{q^{2}-k_{m1}^{2}(\omega_{0})}$,
$k_{m1}(\omega_{0})=\omega_{0}\sqrt{\mu_{0}\varepsilon_{m1}(\omega_{0})}$,
$\varepsilon_{m1}(\omega_{0})=\varepsilon_{0}\varepsilon_{rm1}(\omega_{0})$.

Note that in the limiting case of no material change at the temporal
boundary $t=0$, $\varepsilon_{rm1}(\omega_{0})=\varepsilon_{rm2}(\omega_{0})$,
which leads to $p_{m1}(\omega_{0})=p_{m2}(\omega_{0})$. Then $R_{22}(z^{\prime},s)\rightarrow R_{22}(\omega_{0})=R_{21}(\omega_{0})$
and $T_{12}(z^{\prime},s)\rightarrow T_{12}(\omega_{0})=T_{11}(\omega_{0})$.

\subsubsection{Dispersion Relation}

The dispersion equations $D_{21}=0$ and $D_{22}=0$ lead to singularities in
the Sommerfeld ($q$-) and $s$-planes. Solving these for $q$ we
obtain
\begin{align}
q(\omega_{0})=q_{SPP1} & =k_{0}(\omega_{0})\sqrt{\frac{\varepsilon_{r11}(\omega_{0})\varepsilon_{r21}(\omega_{0})}{\varepsilon_{r11}(\omega_{0})+\varepsilon_{r21}(\omega_{0})}}\label{eq:dispersion q(omega_o)}
\end{align}
for $D_{21}=0$ and
\begin{align}
q(s) & =q_{SPP2}=k_{0s}(s)\sqrt{\frac{\varepsilon_{r12}(s)\varepsilon_{r22}(s)}{\varepsilon_{r12}(s)+\varepsilon_{r22}(s)}},\label{eq:dispersion q(s)}
\end{align}
where $k_{0s}(s)=is\sqrt{\mu_{0}\varepsilon_{0}}$, for $D_{22}=0$.
In particular, $D_{21}=0$ leading to (\ref{eq:dispersion q(omega_o)})
describes the old SPP (before the time boundary) at the source frequency.
After the time boundary, this SPP is not an eigenmode of the new material
system, and, hence, is not a zero of $D_{22}=0$. However, in the
Sommerfeld integral the pole at $q=q_{SPP1}$ is encountered, the
residue of which fixes $q=q_{SPP1}$ in that residue term, independent
of $s$. That term inverse Laplace transforms to a decaying exponential,
since the old SPP cannot propagate in the new material. However, it
cannot simply cease to exist at $t=0$, and so it sheds energy into
radiation and bulk modes. Although $q=q_{SPP1}$ is not an eigenmode
for any $s=i\omega$ with $\omega$ real-valued, the dispersion equation
$D_{22}=0$ can be satisfied at $q=q_{SPP1}$ at a complex value of
$\omega$, the real part being the well-known frequency shift associated
with the time boundary and the imaginary part being the decay constant.
Although not detailed here due to space limitations, a $2nd$-order
Taylor series expansion of $D_{22}=0$ about $s=-i\omega_{0}$ shows
excellent agreement with the well-known frequency shift for the special
case of homogeneous media, and the expected behavior of the decay
constant.

\subsubsection{Fields}

For a vertical unit point dipole current source, $\mathbf{j}_{s}(\mathbf{r})=\widehat{\mathbf{z}}\delta(\mathbf{r}-\mathbf{r}_{0})$,
using $\mathbf{\Upsilon}_{mn}^{e}=\left(k_{mn}^{2}+\mathbf{\nabla}\mathbf{\nabla}\cdot\right)\intop_{V}dV^{\prime}\underline{\mathbf{g}}_{mn}(\mathbf{r},\mathbf{r}^{\prime})\cdot\mathbf{S}_{n}(\mathbf{r}^{\prime})$
and $\mathbf{\Upsilon}_{mn}^{m}=K_{mn}\mathbf{\nabla}\times\intop_{V}dV^{\prime}\underline{\mathbf{g}}_{mn}(\mathbf{r},\mathbf{r}^{\prime})\cdot\mathbf{S}_{n}(\mathbf{r}^{\prime})$,
where $\mathbf{\Upsilon}_{m1}^{e}=\mathbf{E}_{m1}(\mathbf{r})$, $\mathbf{\Upsilon}_{m2}^{e}=\mathbf{E}_{m2}^{\prime}(\mathbf{r},s)$,
$\mathbf{\Upsilon}_{m1}^{m}=\mathbf{H}_{m1}(\mathbf{r})$, $\mathbf{\Upsilon}_{m2}^{m}=\mathbf{H}_{m2}^{\prime}(\mathbf{r},s)$,
$\mathbf{S}_{1}(\mathbf{r}^{\prime})=\nicefrac{\mathbf{j}_{s}(\mathbf{r}^{\prime})}{\left(-i\omega_{0}\varepsilon_{21}\right)}$,
$\mathbf{S}_{2}(\mathbf{r}^{\prime})=\nicefrac{\mathbf{j}_{s}^{\prime}(\mathbf{r}^{\prime},s)}{\left(s\varepsilon_{22}(s)\right)}$,
$K_{m1}=-i\omega_{0}\varepsilon_{m1}$, $K_{m2}=s\varepsilon_{m2}(s)$,
and $\mathbf{j}_{s}^{\prime}(\mathbf{r}^{\prime},s)=A_{2}^{s}(s)\mathbf{j}_{s}(\mathbf{r}^{\prime})$,
we can write the fields for each respective temporal region as
\begin{align}
\mathbf{\Upsilon}_{mn}^{e} & =C_{mn}^{e}\left(\begin{array}{c}
\widehat{\mathbf{z}}g_{zzmn}(\mathbf{r},\mathbf{r}^{\prime})+\frac{1}{k_{mn}^{2}}\left(\widehat{\mathbf{x}}\frac{\partial}{\partial x}\frac{\partial}{\partial z}g_{zzmn}(\mathbf{r},\mathbf{r}^{\prime})+\widehat{\mathbf{y}}\frac{\partial}{\partial y}\frac{\partial}{\partial z}g_{zzmn}(\mathbf{r},\mathbf{r}^{\prime})+\widehat{\mathbf{z}}\frac{\partial^{2}}{\partial z^{2}}g_{zzmn}(\mathbf{r},\mathbf{r}^{\prime})\right)\end{array}\right),\label{eq:electric field}\\
\mathbf{\Upsilon}_{mn}^{m} & =C_{mn}^{m}\left(\widehat{\mathbf{x}}\frac{\partial}{\partial y}g_{zzmn}(\mathbf{r},\mathbf{r}^{\prime})-\widehat{\mathbf{y}}\frac{\partial}{\partial x}g_{zzmn}(\mathbf{r},\mathbf{r}^{\prime})\right),
\end{align}
\end{widetext}where $C_{11}^{e}=i\omega_{0}\mu_{0}\left(\nicefrac{\varepsilon_{11}}{\varepsilon_{21}}\right)$,
$C_{21}^{e}=i\omega_{0}\mu_{0}$, $C_{12}^{e}=A_{2}^{s}(s)\left(-s\mu_{0}\right)\left(\nicefrac{\varepsilon_{12}(s)}{\varepsilon_{22}(s)}\right)$,
$C_{22}^{e}=A_{2}^{s}(s)\left(-s\mu_{0}\right)$, $C_{11}^{m}=\nicefrac{\varepsilon_{11}}{\varepsilon_{21}}$,
$C_{21}^{m}=1$, $C_{12}^{m}=A_{2}^{s}(s)\left(\nicefrac{\varepsilon_{12}(s)}{\varepsilon_{22}(s)}\right)$,
and $C_{22}^{m}=A_{2}^{s}(s)$.

\subsection{Fields in the Time Domain}

We can obtain the fields in the time domain in temporal region 1 using
simply $\mathrm{Re}\left\{ \left(\cdot\right)e^{-i\omega_{0}t}\right\} $
and in temporal region 2 using inverse Laplace transforms.

\section{Results}

In the following we obtain results for material configurations that
give a strong SPP response. As a check on the validity of the model
presented here, we obtained results for the configurations used in
\cite{Jiang1975}, which concerns a dipole source in a spatially-homogeneous
time-changing medium. We found good agreement with the results (Figs.
1-3) in \cite{Jiang1975}, with some small differences attributed
to the different numerical methods used. We also obtained the same
shifted frequency values determined in \cite{Jiang1975}. These comparisons
provided confidence that the model is correct, since here we only
change the Green's function to account for inhomogeneous media and
the presence of SPP-supporting interfaces. We note that the long-time
transient field response approaches the steady-state field after the
temporal-boundary transients have died out, which must occur, and
leads to confidence in the method. Additionally, we observed the effects
of the interference between the forward and backward waves (from the
temporal-boundary transient SPPs), which aligns with the observations
described in \cite{Maslov2018,Raziman}. In that work, a propagating
SPP is the wave incident on a temporal boundary (not the field of
a dipole source as we consider here). We note that determining the
shifted frequency values in this (exact, up to numerical evaluation)
work becomes much more difficult since there are numerous modal temporal
frequencies for the supported SPP modes. Further analysis regarding
this, as well as other areas, e.g., energy conversion at the temporal
boundary (potentially utilizing FDTD simulations), would be beneficial
and a worthwhile pursuit for future work. Here, we focused on first
establishing the necessary framework for further exploration of these
areas, where we emphasis that the checks performed here fully instill
confidence in the validity of the presented framework and model. In
all cases the results are obtained for the configuration in Fig. \ref{fig:Model-for-the-1},
where, as before, we assume nonmagnetic materials, piece-wise homogeneous
layers, and isotropic media.

The permittivity expression used for the dielectric regions (for fields
with time variations of the form $e^{-i\omega t}$) is
\begin{align}
\varepsilon_{r}(\omega) & =\varepsilon_{r}^{\prime}+i\frac{\sigma}{\omega\varepsilon_{0}},\label{eq:permittivity_lossy}
\end{align}
which is the complex permittivity.

The permittivity expression used for the plasma regions is
\begin{equation}
\varepsilon_{r}(\omega)=1-\frac{\omega_{p}^{2}}{\omega^{2}+i\omega\Gamma_{d}},\label{eq:permittivity_Drude}
\end{equation}
which is the Drude dispersion model for the plasma, where $\omega_{p}$
is the plasma frequency (which is a function of the material free
electron density) and $\Gamma_{d}$ is the damping constant (damping
frequency, electron mean collision frequency, i.e., $\Gamma_{d}=\nicefrac{1}{\tau}$,
where $\tau$ is the electron mean collision rate or the electron
momentum scattering time), where, by definition, it accounts for temporally
dispersive materials regardless of whether or not the material is
lossy ($\Gamma_{d}$ accounts for loss, where $\Gamma_{d}=0$ describes
lossless materials).

Then, conforming to the configuration modeled in Fig. \ref{fig:Model-for-the-1},
we use (\ref{eq:permittivity_lossy}) and (\ref{eq:permittivity_Drude})
for the permittivity expressions, where we let $\omega\rightarrow\omega_{0}$
for temporal region 1 and we Laplace transform the expressions ($\omega\rightarrow is$)
for temporal region 2. We consider two cases. One is described in
Table \ref{tab:Relevant-values-used}, with a homogeneous dielectric
space in temporal region 1 and a dielectric-plasma interface in temporal
region 2. We also consider dielectric-plasma interfaces in both temporal
regions, described in Table \ref{tab:Relevant-values-used-1}.
\begin{table}
\centering{}\caption{Permittivity expressions for configuration of a homogeneous dielectric
in temporal region 1 and a dielectric-plasma half-space in temporal
region 2.\label{tab:Relevant-values-used}}
\begin{tabular}[b]{l|>{\raggedright}m{104pt}|>{\centering}m{104pt}|}
\cline{2-3}
 & \centering{}\vspace{1pt}
Temporal Region 1 & \centering{}\vspace{1pt}
Temporal Region 2\tabularnewline
\cline{2-3}
$\hspace*{-14pt}{\color{red}\mathinner{\normalcolor \left.\begin{array}{c}
z\\
\\\\\end{array}\hspace*{-0.15cm}\right\uparrow }\hspace*{-10pt}}$\vspace{-0.6pt}
 & \multirow{2}{104pt}{\centering{}$\begin{array}{c}
\\\varepsilon_{r21}(\omega_{0})=\varepsilon_{r21}^{\prime}+i\frac{\sigma_{21}}{\omega_{0}\varepsilon_{0}}\\
\\\end{array}$\vspace{-20pt}
} & \centering{}$\begin{array}{c}
\\\varepsilon_{r22}(s)=\varepsilon_{r22}^{\prime}+\frac{\sigma_{22}}{s\varepsilon_{0}}\\
\\\end{array}$\vspace{3pt}
\tabularnewline
\cline{3-3}
 &  & \centering{}$\begin{array}{c}
\\\varepsilon_{r12}(s)=1+\frac{\omega_{p12}^{2}}{s^{2}+s\Gamma_{d12}}\\
\\\end{array}$\tabularnewline
\cline{2-3}
\end{tabular}
\end{table}
\begin{table}
\caption{Permittivity expressions for the configuration of a dielectric-plasma
half-space in temporal region 1 and a different dielectric-plasma
half-space in temporal region 2.\label{tab:Relevant-values-used-1}}

\centering{}%
\begin{tabular}[b]{l|>{\centering}m{104pt}|>{\centering}m{104pt}|}
\cline{2-3}
 & \centering{}\vspace{1pt}
Temporal Region 1 & \centering{}\vspace{1pt}
Temporal Region 2\tabularnewline
\cline{2-3}
$\hspace*{-14pt}{\color{red}\mathinner{\normalcolor \left.\begin{array}{c}
z\\
\\\\\end{array}\hspace*{-0.15cm}\right\uparrow }\hspace*{-10pt}}$\vspace{-0.6pt}
 & \centering{}$\begin{array}{c}
\\\varepsilon_{r21}(\omega_{0})=\varepsilon_{r21}^{\prime}+i\frac{\sigma_{21}}{\omega_{0}\varepsilon_{0}}\\
\\\end{array}$\vspace{3pt}
 & \centering{}$\begin{array}{c}
\\\varepsilon_{r22}(s)=\varepsilon_{r22}^{\prime}+\frac{\sigma_{22}}{s\varepsilon_{0}}\\
\\\end{array}$\vspace{3pt}
\tabularnewline
\cline{2-3}
 & \centering{}$\begin{array}{c}
\\\varepsilon_{r11}(\omega_{0})=1-\frac{\omega_{p11}^{2}}{\omega_{0}^{2}+i\omega_{0}\Gamma_{d11}}\\
\\\end{array}$ & \centering{}$\begin{array}{c}
\\\varepsilon_{r12}(s)=1+\frac{\omega_{p12}^{2}}{s^{2}+s\Gamma_{d12}}\\
\\\end{array}$\tabularnewline
\cline{2-3}
\end{tabular}
\end{table}

For the dielectric regions we use $\varepsilon_{r21}^{\prime}=\varepsilon_{r22}^{\prime}=1$
and $\sigma_{21}=\sigma_{22}=0.001\omega_{0}\varepsilon_{0}$. That
is, since we need all materials to be dispersive, i.e., at least slightly
lossy, we will consider a model of air in the limiting low-loss case.
In the case of the plasma regions we will specify the parameters per
each scenario investigated in the applicable results sections. The
source excitation frequency and polarization that we use for all of
the results are $f_{0}=\nicefrac{\omega_{0}}{\left(2\pi\right)}=15$
THz and $\mathbf{j}_{s}=\widehat{\mathbf{z}}j_{zs}$, respectively.

Note that in practice these configurations may be created for experimentation
(practical application) by laser-induced plasma creation for the case
of creating a sudden dense plasma channel in air \cite{MehrpourBernety,Zuo,Lu,Kuo}
(applicable to the configuration described in Table \ref{tab:Relevant-values-used}),
and, for the case of time-varying dielectric-plasma interfaces in
both temporal regions (the configuration described in Table \ref{tab:Relevant-values-used-1}),
the method in \cite{Temnov} can be applied, where the material parameters
of a SPP-supporting platform (at optical frequencies) are modulated
in time by a fast-switching magnetic bias.

\begin{figure}
\begin{centering}
\includegraphics[width=1\columnwidth]{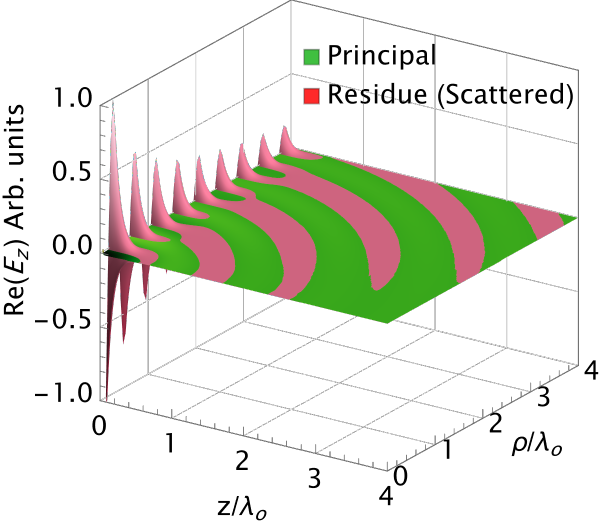}
\par\end{centering}
\caption{Real part of the z-component of each component of the electric field,
showing the principal field (the direct source field in air; obtained
using (\ref{eq:principal Green's function component}) in (\ref{eq:electric field}))
and the scattered field approximated as (strongly dominated by) the
residue contribution \cite{Rothwell2010}, in arbitrary units, as
a function of the observation point height $z$ and the in-plane source-observation
separation distance $\rho=\sqrt{\left(x-x^{\prime}\right)^{2}+\left(y-y^{\prime}\right)^{2}}$,
where the source point height is $z^{\prime}=\nicefrac{\lambda_{0}}{100}$
and the permittivity configuration is an air-plasma half-space, where
the plasma parameters are $\omega_{p12}=1.5\omega_{0}$, $\Gamma_{d12}=0.001\omega_{p12}$;
$\varepsilon_{r12}(\omega_{0})=-1.25+i0.003$ (we are using the configuration
in Table \ref{tab:Relevant-values-used}, albeit, we are only using
the configuration in temporal region 2 on its own (no time boundary,
no temporal region 1) for the steady state case).\label{fig:Plot-of-the-1}}
\end{figure}

\begin{figure}
\subfloat{\centering{}\stackinset{l}{0pt}{b}{79pt}{(a)}{}\includegraphics[width=0.5\columnwidth,height=29mm]{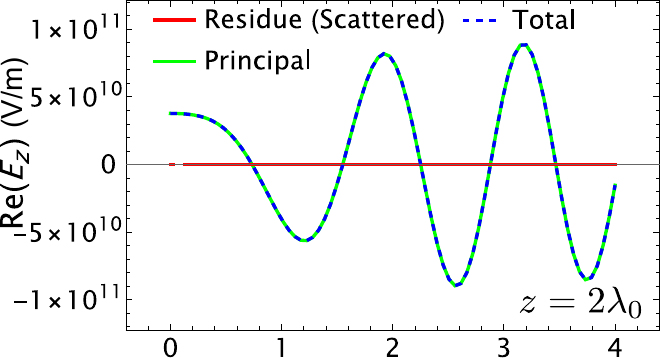}}\,\subfloat{\centering{}\stackinset{l}{0pt}{b}{79pt}{(b)}{}\includegraphics[width=0.48\columnwidth,height=29mm]{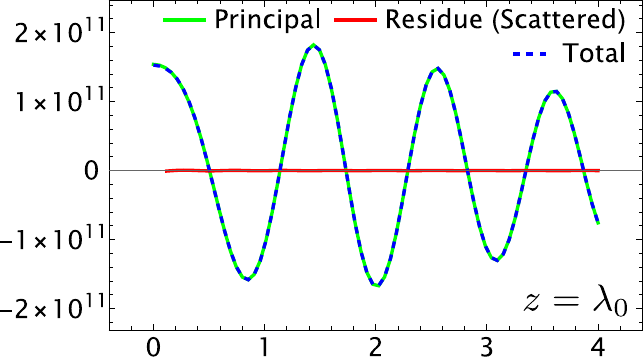}}\vspace{-0.35cm}
\subfloat{\centering{}\stackinset{l}{0pt}{b}{79pt}{(c)}{}\includegraphics[width=0.5\columnwidth,height=29mm]{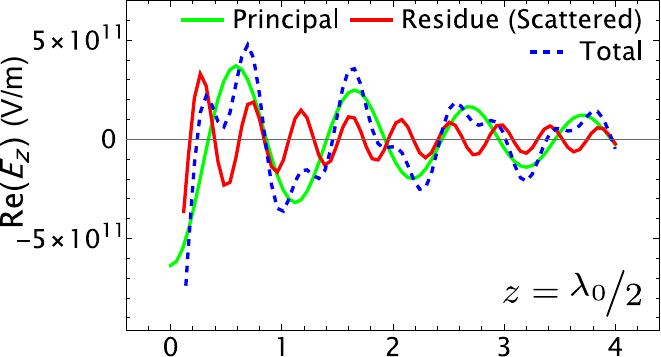}}\,\subfloat{\centering{}\stackinset{l}{0pt}{b}{79pt}{(d)}{}\includegraphics[width=0.48\columnwidth,height=29mm]{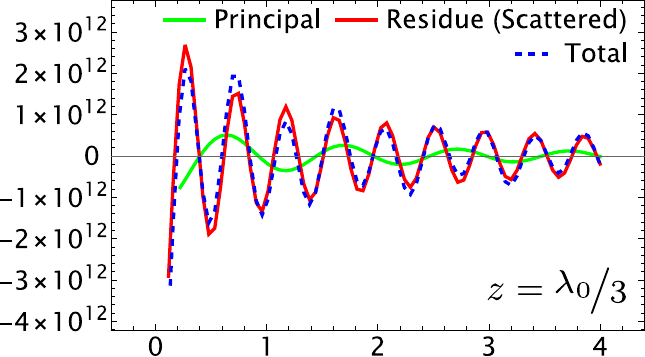}}\vspace{-0.4cm}
\subfloat{\centering{}\stackinset{l}{0pt}{b}{96pt}{(e)}{}\includegraphics[width=0.5\columnwidth,height=34mm]{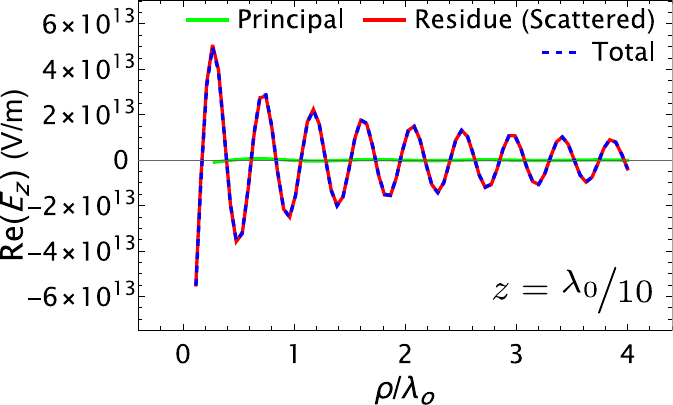}}\,\subfloat{\centering{}\stackinset{l}{0pt}{b}{96pt}{(f)}{}\includegraphics[width=0.48\columnwidth,height=34mm]{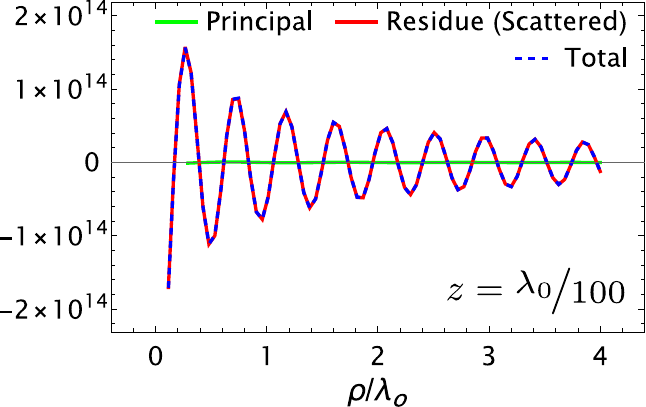}}\caption{Real part of the z-component of each respective sinusoidal steady-state
electric field, showing the principal field (the direct source field
in air; obtained using (\ref{eq:principal Green's function component})
in (\ref{eq:electric field})), the scattered field approximated as
(strongly dominated by) the residue contribution \cite{Rothwell2010},
and the total field, where the total field is the principal field
plus the approximate scattered field, as a function of the observation
point height $z$ above the interface: (a) $z=2\lambda_{0}$, (b)
$z=\lambda_{0}$, (c) $z=\nicefrac{\lambda_{0}}{2}$, (d) $z=\nicefrac{\lambda_{0}}{3}$,
(e) $z=\nicefrac{\lambda_{0}}{10}$, (f) $z=\nicefrac{\lambda_{0}}{100}$,
where the in-plane source-observation separation distance is $\rho=2.2\lambda_{0}$,
the source point height is $z^{\prime}=\nicefrac{\lambda_{0}}{100}$,
and the permittivity configuration is an air-plasma half-space, where
the plasma is $\omega_{p12}=1.5\omega_{0}$, $\Gamma_{d12}=0.001\omega_{p12}$,
$\varepsilon_{r12}=-1.25+i0.003$ (we are using the configuration
in Table \ref{tab:Relevant-values-used}, albeit, we are only using
the configuration in temporal region 2 on its own (no time boundary,
no temporal region 1) for the steady state case) for all cases.\label{fig:Plot-of-the}}
\end{figure}

In Figs. \ref{fig:Plot-of-the-1} and \ref{fig:Plot-of-the}, we first
plot the sinusoidal steady state response in the frequency domain
for an air-plasma half-space (no time boundary) to gain some insight
into SPP behavior. For these plots we use the principal field (the
direct source field in air; obtained using (\ref{eq:principal Green's function component})
in (\ref{eq:electric field})) and the scattered field approximated
as (strongly dominated by) the residue contribution \cite{Rothwell2010},
where the total field is the principal field plus the approximate
scattered field.

Figure \ref{fig:Plot-of-the-1} shows a typical SPP response, where
we see the SPP characteristic of confinement to the surface (exponential
decay away from the surface in the $z$-direction) \cite{Maier2010}.
Additionally, we see the effects of decay with increasing dipole-interface
separation, and loss as the SPPs propagate along the surface, in the
$\rho$-direction. We also see these effects, for a fixed $\rho$,
in Fig. \ref{fig:Plot-of-the}, where we can see the transition from
the propagation in air to SPP propagation as we approach the surface
from above, in the $z$-direction. At large distances away from the
interface, the principal and total fields are approximately the same
(the principal field is the dominant contribution to the total field),
but closer to the interface the SPP (contribution from residue) dominates
the field.

\subsection{Dynamic SPP Formation}

\begin{figure}
\subfloat{\centering{}\stackinset{l}{0pt}{b}{40pt}{(a)}{}%
\begin{minipage}[c][40mm]{49mm}%
\begin{center}
\includegraphics[width=49mm,height=40mm]{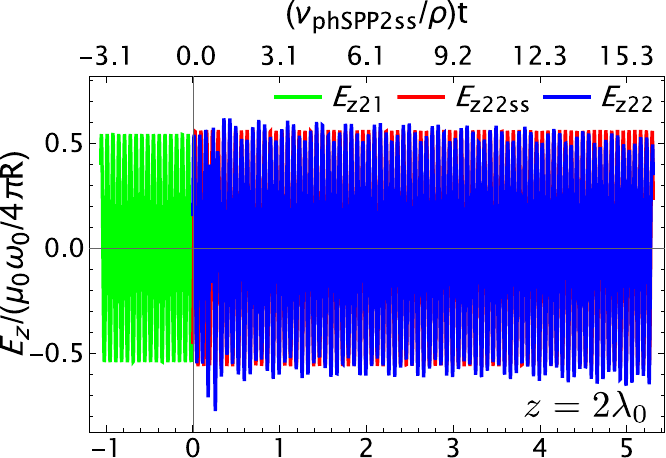}
\par\end{center}%
\end{minipage}\,%
\begin{minipage}[c][30mm]{37mm}%
\vspace{0.8cm}

\begin{center}
\includegraphics[width=37mm,height=30mm]{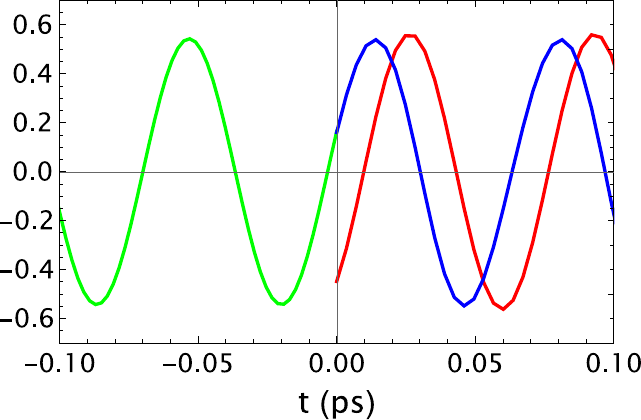}
\par\end{center}%
\end{minipage}}

\vspace{-0.2cm}

\subfloat{\centering{}\stackinset{l}{0pt}{b}{45pt}{(b)}{}%
\begin{minipage}[c][36mm]{49mm}%
\begin{center}
\includegraphics[width=49mm,height=36mm]{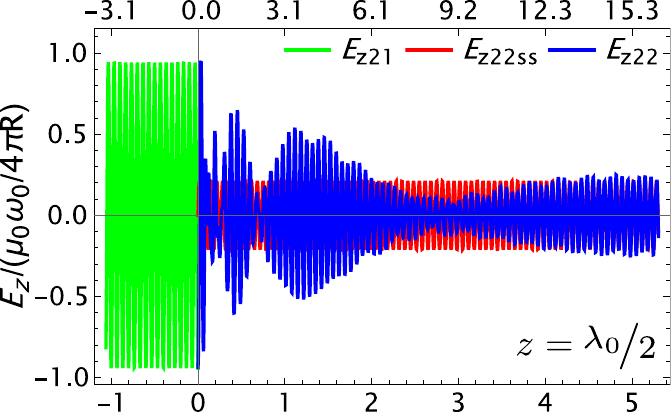}
\par\end{center}%
\end{minipage}\,%
\begin{minipage}[c][30mm]{37mm}%
\vspace{0.7cm}

\begin{center}
\includegraphics[width=37mm,height=30mm]{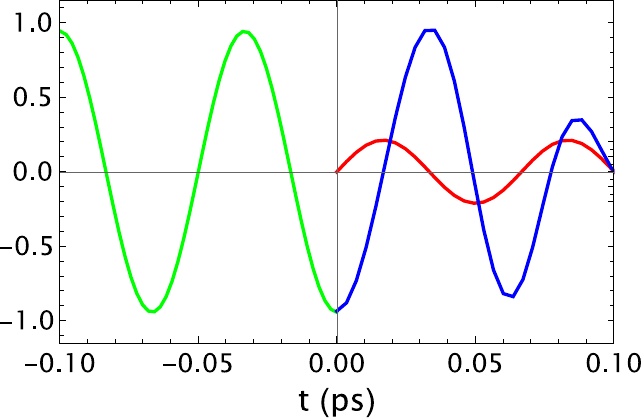}
\par\end{center}%
\end{minipage}}

\vspace{0.2cm}

\noindent\hspace{0.2cm}\subfloat{\centering{}\stackinset{l}{-2.5pt}{b}{90pt}{(c)}{}\includegraphics[width=44mm,height=36mm]{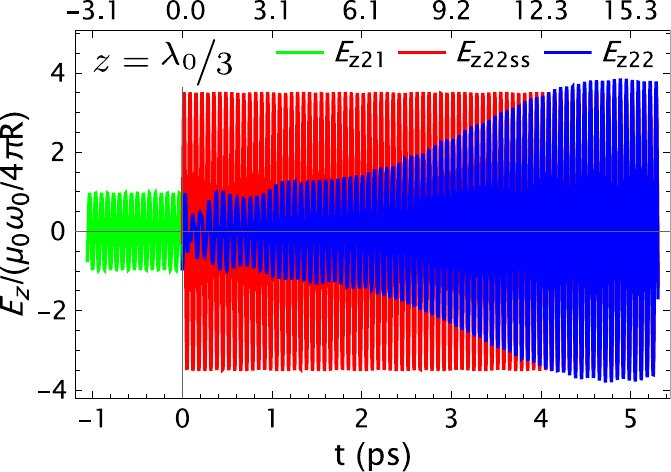}}\hspace{0.1cm}\subfloat{\centering{}\stackinset{l}{-2.5pt}{b}{90pt}{(d)}{}\includegraphics[width=40mm,height=36mm]{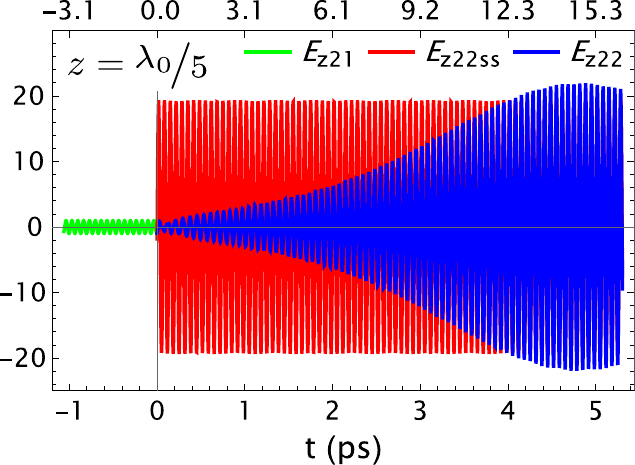}}

\caption{Temporal-boundary induced SPP formation. Real part of the electric
field versus time for the steady state response in temporal region
1 ($E_{z21}$), the steady state response in temporal region 2 ($E_{z22ss}$),
and the transient electric field in temporal region 2 ($E_{z22}$),
as a function of the observation point height $z$ above the interface:
(a) $z=2\lambda_{0}$, with additional plots, zoomed in around $t=0$
(temporal boundary), (b) $z=\nicefrac{\lambda_{0}}{2}$, with additional
plots, zoomed in around $t=0$ (temporal boundary), (c) $z=\nicefrac{\lambda_{0}}{3}$,
and (d) $z=\nicefrac{\lambda_{0}}{5}$, where the in-plane source-observation
separation distance is $\rho=2.2\lambda_{0}$, the source point height
is $z^{\prime}=\nicefrac{\lambda_{0}}{100}$, and we use the configuration
in Table \ref{tab:Relevant-values-used}, where the temporal region
2 plasma parameters are $\omega_{p12}=1.5\omega_{0}$, $\Gamma_{d12}=0.001\omega_{p12}$,
$\varepsilon_{r12}=-1.25+i0.003$ for all cases. We can see that the
fields are continuous (the field in temporal region 1 (green) and
the field in temporal region 2 (blue), which includes the transient
response) at the temporal boundary $t=0$. Note that $R=\left|\mathbf{r}-\mathbf{r}^{\prime}\right|=\sqrt{\rho^{2}+\left(z-z^{\prime}\right)^{2}}$,
where $\rho=\sqrt{\left(x-x^{\prime}\right)^{2}+\left(y-y^{\prime}\right)^{2}}$.\label{fig:Plot-of-the-2}}
\end{figure}

We now introduce a temporal interface and examine the SPP formation
at a time-boundary, and the field response versus time (normalized
time defined as $\left(\nicefrac{\nu_{phSPP2ss}}{\rho}\right)t$,
where $\nu_{phSPP2ss}=\nicefrac{\omega_{0}}{\mathrm{Re}\left\{ q_{SPP2ss}\left(\omega_{0}\right)\right\} }$
is the phase velocity for the steady state SPP \cite{Ioannidis,Ziyatkhan,Gric}
in temporal region 2), at various heights above the interface, to
demonstrate dynamic SPP formation. We use the configuration in Table
\ref{tab:Relevant-values-used}, where the temporal region 2 plasma
parameters are $\omega_{p12}=1.5\omega_{0}$, $\Gamma_{d12}=0.001\omega_{p12}$.
In Fig. 4, before the time-boundary we plot the electric field in
temporal region 1, $E_{z21}$, which is simply the usual sinusoidal
response to a time-harmonic source. Starting at the time-boundary,
when the media configuration changes from a homogeneous dielectric
space to a dielectric-plasma interface (Table \ref{tab:Relevant-values-used}),
at which point an SPP can form, we have $E_{z22}$, forming a transient
response. For reference, we also plot the sinusoidal steady-state
response for the temporal region 2 media configuration, $E_{z22ss}$,
which the transient field must tend to (for $t\gg\nicefrac{\nu_{phSPP2ss}}{\rho}$)
if the response is stable. In other words, $E_{z22ss}$ is the value
for the field response resulting from the temporal region 2 media
configuration (the field response for that media configuration on
its own (no time boundary, no temporal region 1; i.e., it's as if
the media was always there)), which the transient field must tend
to once the system settles down (this provides a further check on
the validity of the model and associated results). The phase velocity
referenced here is determined as described previously in this section.
The fields are plotted for different observation point heights, compared
to the wavelength $\lambda_{0}$, specified in the caption of Fig.
\ref{fig:Plot-of-the-2}. We obtain the fields, using the total Green's
function (the principal Green's function obtained from (\ref{eq:principal Green's function component})
plus the scattered Green's function obtained from (\ref{eq:scattered Green's function component}))
in (\ref{eq:electric field}), where the scattered Green's function
is not approximated as its residue; it is obtained from the formal
integration. We can see that at a height greater than a wavelength
$\lambda_{0}$ the field in air (i.e., coupling to a continuum, as
opposed to the SPP, which is a guided mode) dominates the total field
response in temporal region 2, however, for heights less than a wavelength
$\lambda_{0}$, the SPP starts to contribute to the total field. We
can also see that the fields are continuous (the field in temporal
region 1 (green) and the field in temporal region 2 (blue), which
includes the transient response) at the temporal boundary $t=0$.
 As a whole, these plots demonstrate the dynamic SPP formation as
the media is suddenly changed from air to an air-plasma half-space.

\subsection{SPP Transient Period and Time to Steady State}

\begin{figure}
\noindent\subfloat{\centering{}\stackinset{l}{0pt}{b}{200pt}{(a)}{}%
\begin{minipage}[b][76mm][c]{43mm}%
\begin{center}
\includegraphics[width=43mm,height=40mm]{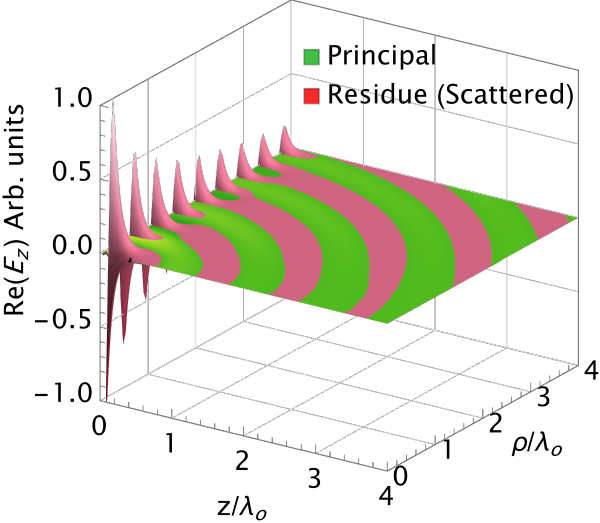}
\par\end{center}
\vspace{-0.3cm}

\begin{center}
\includegraphics[width=43mm,height=36mm]{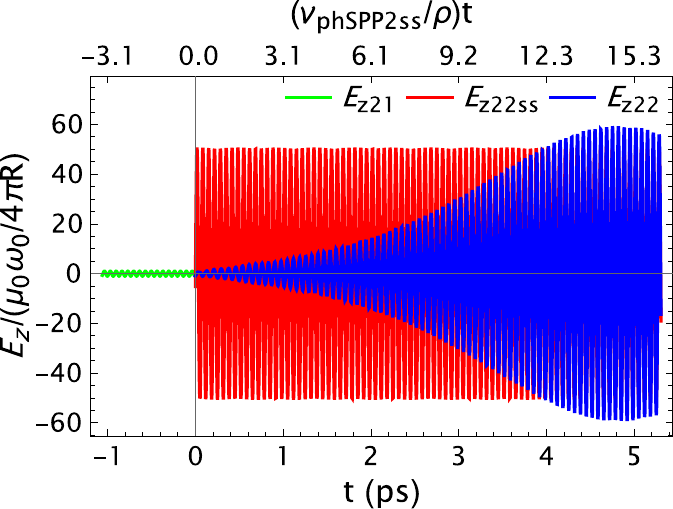}
\par\end{center}%
\end{minipage}}\,\subfloat{\centering{}\stackinset{l}{0pt}{b}{200pt}{(b)}{}%
\begin{minipage}[b][76mm][c]{43mm}%
\begin{center}
\includegraphics[width=43mm,height=40mm]{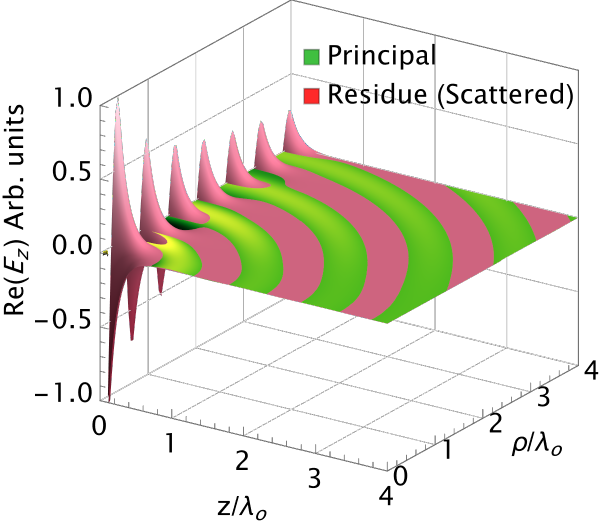}
\par\end{center}
\vspace{-0.3cm}

\begin{center}
\includegraphics[width=42mm,height=36mm]{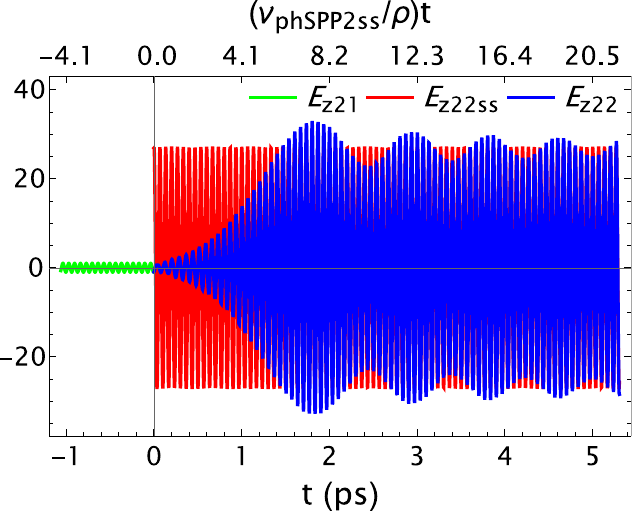}
\par\end{center}%
\end{minipage}}

\vspace{0cm}

\noindent\subfloat{\centering{}\stackinset{l}{0pt}{b}{200pt}{(c)}{}%
\begin{minipage}[b][76mm][c]{43mm}%
\begin{center}
\includegraphics[width=43mm,height=40mm]{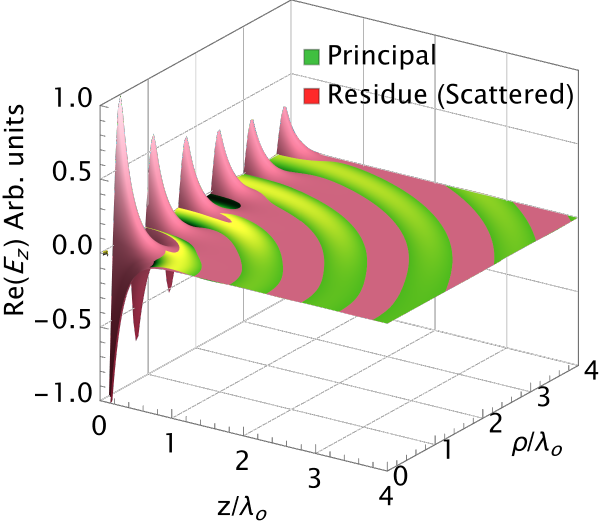}
\par\end{center}
\vspace{-0.3cm}

\begin{center}
\includegraphics[width=43mm,height=36mm]{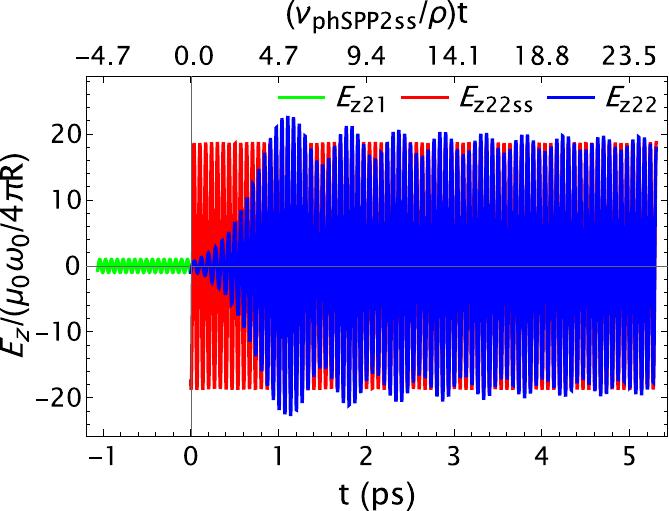}
\par\end{center}%
\end{minipage}}\,\subfloat{\centering{}\stackinset{l}{0pt}{b}{200pt}{(d)}{}%
\begin{minipage}[b][76mm][c]{43mm}%
\begin{center}
\includegraphics[width=43mm,height=40mm]{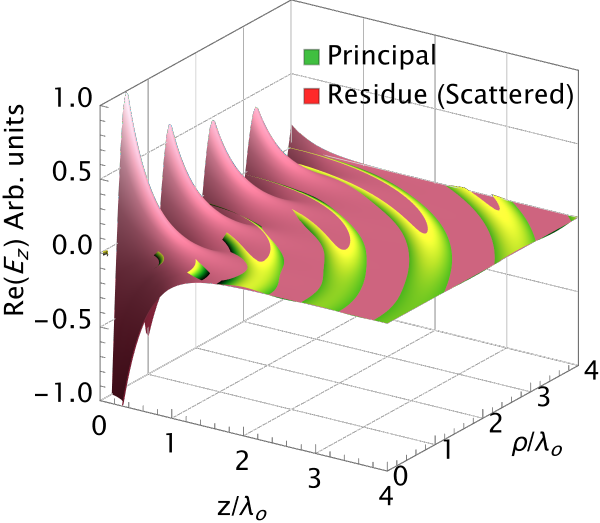}
\par\end{center}
\vspace{-0.3cm}

\begin{center}
\includegraphics[width=42mm,height=36mm]{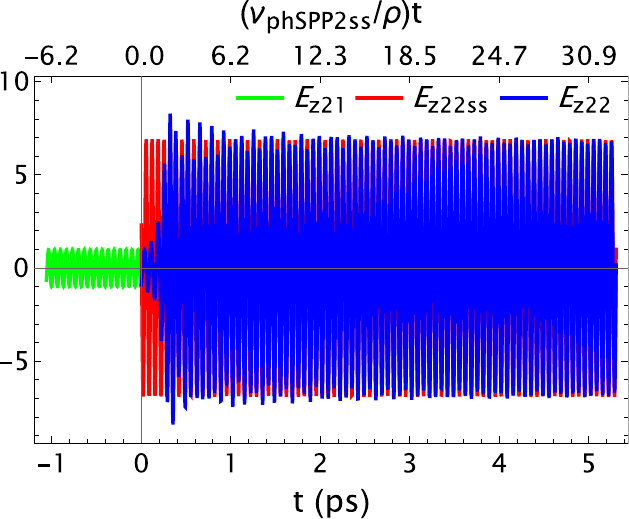}
\par\end{center}%
\end{minipage}}

\caption{Plots similar to the plots in Figs. \ref{fig:Plot-of-the-1} (the
3D plots) and \ref{fig:Plot-of-the-2} (the line plots), where now
we consider the effect of different permittivity configurations (the
plasma material) in temporal region 2. We use the configuration in
Table \ref{tab:Relevant-values-used}, where we increase the plasma
frequency $\omega_{p12}$: (a) $\omega_{p12}=1.5\omega_{0}$, $\varepsilon_{r12}=-1.25+i0.003$,
(b) $\omega_{p12}=1.6\omega_{0}$, $\varepsilon_{r12}=-1.56+i0.004$,
(c) $\omega_{p12}=1.7\omega_{0}$, $\varepsilon_{r12}=-1.89+i0.005$,
and (d) $\omega_{p12}=2.5\omega_{0}$, $\varepsilon_{r12}=-5.25+i0.016$.
In all cases, the source excitation frequency is $f_{0}=\nicefrac{\omega_{0}}{\left(2\pi\right)}=15$
THz, the damping frequency in the plasma material in temporal region
2 is $\Gamma_{d12}=0.001\omega_{p12}$, the in-plane source-observation
separation distance is $\rho=2.2\lambda_{0}$, the source point height
is $z^{\prime}=\nicefrac{\lambda_{0}}{15}$, and the observation point
height is $z=\nicefrac{\lambda_{0}}{15}$.\label{fig:Plot-of-the-3}}
\end{figure}

We now consider the effect of different permittivity configurations
(a change to the plasma material) in temporal region 2 on the SPP
transient period and time to steady state, where a transient period
after the temporal boundary occurs due to the causal response of the
system, i.e., it takes time for the SPP to form after the medium is
suddenly changed since the material response (electrons) cannot change
instantaneously due to their mass. In this case, we use a source point
height of $z^{\prime}=\nicefrac{\lambda_{0}}{15}$ and we use the
configuration in Table \ref{tab:Relevant-values-used}. In Fig. \ref{fig:Plot-of-the-3},
where we have plots similar to the plots in Figs. \ref{fig:Plot-of-the-1}
(the 3D plots) and \ref{fig:Plot-of-the-2} (the line plots), we can
see that as we increase the plasma frequency the plasma permittivity
becomes more negative (more like a metal), and the electrons in the
plasma can more quickly respond to the excitation screening the field,
resulting in a less confined SPP and longer SPP wavelengths. Furthermore,
more reflection back into the air region can occur. Therefore, as
the plasma permittivity becomes more negative, the transient response
becomes faster, the SPP amplitude decreases (as we go from subfigure
(a) to (d) in Fig. \ref{fig:Plot-of-the-3} we can see that the overall
amplitude decreases as a result of more radiation into the air region),
and steady state is approached much sooner. We also note that amplitude
modulation can be seen occurring due to the interference between the
forward and backward waves (from the SPPs and direct radiation) in
temporal region 2, which have shifted frequencies relative to the
waves in temporal region 1, that result due to the momentum conservation
at the temporal boundary. 
 Although not shown, as we increase the in-plane observation-source
separation, the transient period is longer and it takes longer to
approach steady state, since it takes more time for the SPP to travel
farther; the system response time is longer.

Additionally, considering the effect of additional loss in the plasma
material in temporal region 2 on the SPP response, at a fixed $\rho$,
we observe that as we increase the loss in the plasma material the
SPP quickly dampens out, which results in a shorter transient period
and time to steady state.

\subsection{Interference of Slow and Fast Propagating SPPs}

Here, we consider the interference of slow and fast propagating SPPs,
using the configuration in Table \ref{tab:Relevant-values-used-1}.
The dipole source exists in both temporal regions, as before. In temporal
region 1 it excites an SPP that propagates through the temporal interface
(that SPP, however, is no longer excited by the source once the interface
has changed at the temporal boundary). In temporal region 2 the source
excites a different SPP, according to the new material configuration.
These two SPPs can interfere until the SPP originally excited in temporal
region 1 dissipates. We use the SPP phase velocity to determine the
SPP arrival time to the observation point, i.e., the time it takes
to travel the in-plane source-observation separation distance, where
$\nu_{phSPP1}=\nicefrac{\omega_{0}}{\mathrm{Re}\left\{ q_{SPP1}\left(\omega_{0}\right)\right\} }$
and $\nu_{phSPP2ss}=\nicefrac{\omega_{0}}{\mathrm{Re}\left\{ q_{SPP2ss}\left(\omega_{0}\right)\right\} }$.
We can then determine the arrival time for SPPs created in temporal
region 1 and the SPPs created in temporal region 2, starting from
the material time change, i.e., $t=0$, as $t_{SPP1}=\nicefrac{\rho}{\nu_{phSPP1}}$
and $t_{SPP2}=\nicefrac{\rho}{\nu_{phSPP2ss}}$, respectively.

We consider two completely separate permittivity configuration scenarios,
with an in-plane source-observation separation distance of $\rho=6\lambda_{0}$,
where $\lambda_{0}=\nicefrac{2\pi}{k_{0}}\approx20$ $\mu$m is vacuum
wavelength. The first scenario is (a): a slow propagating SPP in temporal
region 1, and a fast propagating SPP in temporal region 2. We then
swap the configurations for temporal region 1 and 2, to see the opposite
SPP response, where we then have (b): a fast propagating SPP in temporal
region 1, and a slow propagating SPP in temporal region 2. Note that
(a) and (b) are completely separate scenarios, where, in general,
a faster propagating SPP corresponds to the plasma with a larger plasma
frequency.

\begin{figure}
\subfloat{\centering{}\label{fig:spp interference_a}\stackinset{l}{0pt}{b}{35pt}{(a)}{}%
\begin{minipage}[c][36mm]{86mm}%
\begin{center}
\includegraphics[width=44mm,height=36mm]{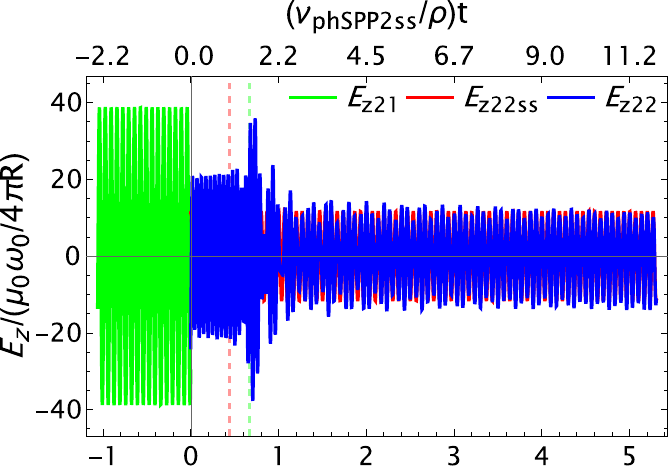}\,\includegraphics[width=42mm,height=30.4mm]{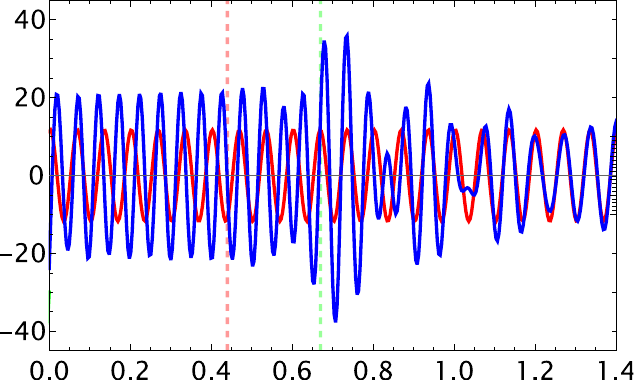}
\par\end{center}%
\end{minipage}}

\vspace{-0.2cm}

\subfloat{\centering{}\label{fig:spp interference_b}\stackinset{l}{0pt}{b}{50pt}{(b)}{}%
\begin{minipage}[c][36mm]{86mm}%
\begin{center}
\includegraphics[width=44mm,height=36mm]{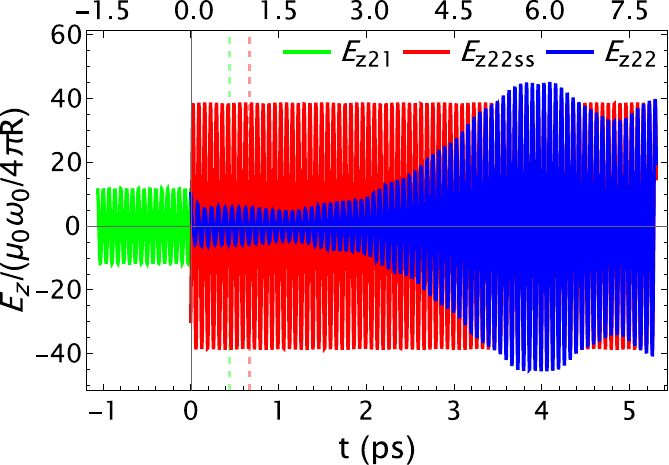}\,\includegraphics[width=42mm,height=33.8mm]{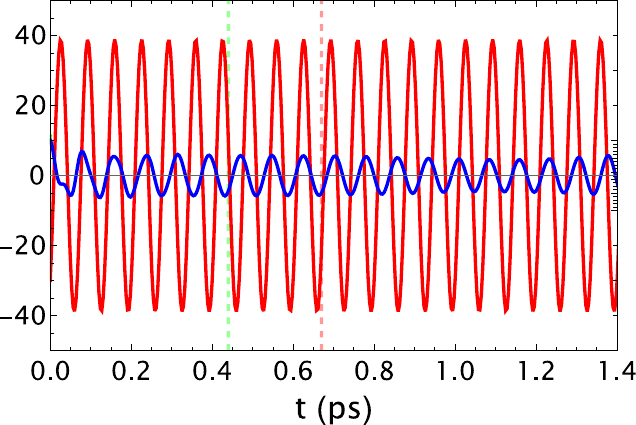}
\par\end{center}%
\end{minipage}}

\caption{Real part of the electric field versus time plots regarding interference
of slow and fast propagating SPPs. We use the configuration in Table
\ref{tab:Relevant-values-used-1}, where the permittivity configuration
for the plasma in temporal region 1 and 2 is: (a) slow propagating
SPP: $\omega_{p11}=1.6\omega_{0}$, $\Gamma_{d11}=0.001\omega_{p11}$,
$\varepsilon_{r11}=-1.56+i0.004$ and fast propagating SPP: $\omega_{p12}=2.5\omega_{0}$,
$\Gamma_{d12}=0.001\omega_{p12}$, $\varepsilon_{r12}=-5.25+i0.016$,
respectively, and (b) fast propagating SPP: $\omega_{p11}=2.5\omega_{0}$,
$\Gamma_{d11}=0.001\omega_{p11}$, $\varepsilon_{r11}=-5.25+i0.016$
and slow propagating SPP: $\omega_{p12}=1.6\omega_{0}$, $\Gamma_{d12}=0.001\omega_{p12}$,
$\varepsilon_{r12}=-1.56+i0.004$, respectively, where the additional
plots are zoomed in around the time range where the slow and fast
propagating SPPs are potentially able to interfere given their arrival
time (green dashed line is for the SPP from temporal region 1, red
dashed line is for the SPP formed in temporal region 2) to the observation
point. In all cases, the in-plane source-observation separation distance
is $\rho=6\lambda_{0}$, the source point height is $z^{\prime}=\nicefrac{\lambda_{0}}{15}$,
and the observation point height is $z=\nicefrac{\lambda_{0}}{15}$.\label{fig:Plot-of-the-3-2-2}}
\end{figure}

Then, for the corresponding SPP phase velocities of $\nu_{phSPP1}=1.8\times10^{8}$
and $\nu_{phSPP2ss}=2.7\times10^{8}$, the arrival times for scenario
(a) are $t_{SPP1_{a}}=0.67$ ps and $t_{SPP2_{a}}=0.44$ ps, and for
scenario (b) ($\nu_{phSPP1}=2.7\times10^{8}$ and $\nu_{phSPP2ss}=1.8\times10^{8}$),
they are $t_{SPP1_{b}}=0.44$ ps and $t_{SPP2_{b}}=0.67$ ps, where
the a and b subscripts correspond to the (a) and (b) scenarios that
we are investigating, respectively. Those arrival times are shown
in the plots in Fig. \ref{fig:Plot-of-the-3-2-2} (the green and red
dashed lines). Note that when we change the media (the plasma) at
$t=0$, the excitation of the SPPs for the plasma in temporal region
1 no longer occurs, instead the excitation of the SPPs for the plasma
in temporal region 2 begins and continues. Since the source remains
before and after the time change, any direct radiation from that will
continue. In all cases it will take time for the fields to reach and
pass the observation point. In the case of the SPP fields, the time
it takes will depend on the SPP phase velocity.

Therefore, the SPP from temporal region 1, which is present at the
observation point before the time change, will continue to be present
at (i.e., propagating past) the observation point until the last part
of the wave, which ceases to be excited after $t=0$, reaches (passes
by) the observation point, after which it will cease to exist (eventually
dying out). The SPP created in temporal region 2, starting at the
time change at $t=0$, will also take time to reach the observation
point, where it will have its own arrival time depending on its phase
velocity, after which it will continue to be excited by the source
and propagate. With all of this taken into consideration we can investigate
the interference of slow and fast propagating SPPs based on their
arrival times.

In the case of a slow propagating SPP from temporal region 1 and a
fast propagating SPP in temporal region 2 (Fig. \ref{fig:spp interference_a}),
the slow propagating SPP (from temporal region 1) has been continuously
passing by the observation point (has always been there). It takes
a relatively long time to stop passing the observation point (0.67
ps). Meanwhile, starting at 0.44 ps, the fast propagating SPP, created
(starting) at $t=0$, reaches the observation point, and since the
slow propagating SPP is still there, the two interfere. After 0.67
ps the slow propagating SPP is gone, and interference ends, which
is what we see at 0.67 ps (the green dashed line) in the plot in Fig.
\ref{fig:spp interference_a}. Given this we see that we can interfere
a slow and fast propagating SPP in a known time frame to induce constructive
interference between the SPPs and achieve a larger amplitude.

We also see interference occurring before this time, after the fast
propagating SPP (from temporal region 2) arrival time and before it.
This interference is occurring between the forward and backward waves
(from the SPPs and direct radiation), which have shifted frequencies
relative to the waves in temporal region 1, that result due to the
momentum conservation at the temporal boundary. Up to this point we
haven't discussed this much, however, we can see it occurring in some
of the other previous plots, where it is more pronounced, e.g., in
Fig. \ref{fig:Plot-of-the-2}. Based on these results we can see that
it is possible to enhance the SPP field intensity by interfering slow
and fast propagating SPPs.

In Fig. \ref{fig:spp interference_b}, the case of a fast propagating
SPP from temporal region 1 and a slow propagating SPP in temporal
region 2, the fast propagating SPP (from temporal region 1) stops
passing the observation point after 0.44 ps, before the slow propagating
SPP, created (starting) at $t=0$, reaches the observation point (after
0.67 ps). Therefore, the SPPs are unable to interfere with each other,
which is what we see (there is no interference) around that time frame
in the plot in Fig. \ref{fig:spp interference_b}.

\section{Conclusion}

We established the theoretical framework for the specific configuration
of a time-varying media system that supports SPPs excited by a dipole
excitation (electric dipole point source), where the media configuration
suddenly changes at a temporal boundary. Such a system allows for
the analysis of the interaction among dipole excitations, where we
have the ability to consider the interactions at the moment of SPP
creation. Using this framework, we then demonstrated dynamic SPP formation
and interference of slow and fast propagating SPPs, for the case of
isotropic media and one temporal boundary. This provided insight into
how SPPs respond in a time-varying media system and, in the case of
the interference of slow and fast propagating SPPs, demonstrated that
we can induce some constructive interference between SPPs in a known
time frame. There are numerous avenues that warrant further research
in the area of time-varying media, where further work is needed to
incorporate the case of anisotropic media and multi-temporal boundaries
into the framework already established here for this type of system.

\bibliographystyle{IEEEtran}
\bibliography{IEEE_TAP_2024_paper_berre_final}

\end{document}